\documentclass[sigconf,screen]{acmart}

\makeatletter
\let\@fnsymbol@original\@fnsymbol
\renewcommand{\@fnsymbol}[1]{%
   \ensuremath{%
      \ifcase#1% 0
      \or \dagger       % 1 (原本是 *, 现在改为 †) ==> Equal contribution
      \or *             % 2 (原本是 †, 现在改为 *) ==> Corresponding author
      \or \ddagger      % 3
      \or \mathsection  % 4
      \or \mathparagraph% 5
      \or \|            % 6
      \or **            % 7
      \or \dagger\dagger% 8
      \or \ddagger\ddagger% 9
      \else\@ctrerr\fi}}
\makeatother

\usepackage{natbib}
\usepackage{pifont}       % \ding{xx}
\usepackage{bbding}       % \Checkmark,\XSolid,... (需要和pifont宏包共同使用)
\usepackage{fontawesome}  % \faCheck,\faTimes
 \usepackage[table]{xcolor}
\usepackage{colortbl}       % for cell color
\usepackage{graphicx}   % 用于插入图像
\usepackage{caption}     % 用于处理标题
\usepackage{subcaption}  % 用于创建子图
\usepackage{mathtools}
\usepackage{enumitem}

\usepackage{wrapfig} % 确保在导言区加载这个宏包
\usepackage[utf8]{inputenc} % allow utf-8 input
\usepackage[T1]{fontenc}    % use 8-bit T1 fonts
\usepackage{xcolor}         % colors
\definecolor{darkgreen}{rgb}{0.0, 0.5, 0.0}
\definecolor{lightblue}{rgb}{0.867, 0.922, 0.969}
\usepackage{array}          % for table
\usepackage{colortbl}       % for table
\usepackage{makecell}       % for table
\usepackage{tabularx}       % for table
\usepackage{booktabs}       % for tables
\usepackage{bigstrut}       % for bigstrut
\usepackage{multirow}       % for multirow
\usepackage{colortbl}       % for cell color
\usepackage{nicefrac}       % compact symbols for 1/2, etc.
\usepackage{microtype}      % microtypography

\usepackage{algorithm}      % for algorithm
\usepackage{algorithmicx}   % for algorithm
\usepackage{algpseudocode}  % for algorithm

\usepackage{thm-restate}    % for restatable theorems
\usepackage{amsfonts}       % blackboard math symbols

\usepackage{amssymb}        % for math symbols
\usepackage{amsthm}         % for theorems

\numberwithin{equation}{section}    % number equations within sections
\numberwithin{algorithm}{section}   % number algorithms within sections
\numberwithin{table}{section}       % number tables within sections

\definecolor{linkdarkblue}{rgb}{0, 0.08, 0.45}    % used for hyperlinks
\usepackage{cleveref}

\usepackage{hhline} % for hline not coverred by color cell

\crefname{appendix}{Appendix}{Appendix}

\newcommand{\lossName}{Talos}

\newcommand{\loss}{\mathcal{L}}
\newcommand{\eg}{\emph{e.g., }}         % for example
\newcommand{\ie}{\emph{i.e., }}         % that is
\newcommand{\wrt}{w.r.t. }
\newcommand{\etc}{\text{etc}}           % and so on
\newcommand{\iid}{\text{i.i.d. }}       % independent and identically distributed
\newcommand{\cf}{\emph{cf. }}
\newcommand{\posU}{\mathcal{P}_u}
\newcommand{\negU}{\mathcal{N}_u}
\newcommand{\itemU}{\mathcal{I}}
\newcommand{\sui}{s_{ui}}
\newcommand{\suj}{s_{uj}}
\newcommand{\margin}{\beta_u^k}

\newcommand{\sumPosU}{\sum\limits_{i\in\posU}}
\newcommand{\sumNegU}{\sum\limits_{j\in\negU}}
\newcommand{\userSet}{\mathcal{U}}
\newcommand{\itemSet}{\mathcal{I}}
\newcommand{\obSet}{\mathcal{D}}

\newcommand{\E}{\mathbb{E}}
\newcommand{\I}{\mathbb{I}}
\newcommand{\topk}{Top-$K$ }

\newcommand{\diff}{\mathrm{d}}          % differential
\newcommand{\cdf}{\text{c.d.f. }}       % cumulative distribution function
\newcommand{\assCons}{d}
\newcommand{\tauQR}{\kappa}

\newcommand{\lossA}{\loss_{\text{QR-2}}}
\newcommand{\lossB}{\loss_{\text{QR-S}}}
\newcommand{\sampledNegative}{G_u}

\def\eq#1{Eq.(#1)}
\definecolor{skyblue}{rgb}{0.53, 0.81, 0.98}

\definecolor{customred}{RGB}{192,0,0} % #C00000
\definecolor{lightblue}{RGB}{238,245,252}  % #EEF5FC
\definecolor{customblue}{RGB}{31,78,120}    % 深蓝色 #1F4E78

\AtBeginDocument{%
}

\copyrightyear{2026}
\acmYear{2026}
\setcopyright{cc}
\acmConference[WWW '26]{Proceedings of the ACM Web Conference 2026}{April 13--17, 2026}{Dubai, United Arab Emirates}
\acmBooktitle{Proceedings of the ACM Web Conference 2026 (WWW '26), April 13--17, 2026, Dubai, United Arab Emirates}
\acmPrice{}

\begin{document}

\title{\lossName: Optimizing Top-$K$ Accuracy in Recommender Systems}

\author{Shengjia Zhang}
    \orcid{0009-0004-0209-2276}
    \affiliation{
        \institution{Zhejiang University}
        \city{Hangzhou}
        \country{China}
    }
    \email{shengjia.zhang@zju.edu.cn}
\authornote{Equal contribution.}
\authornotemark[3]
\authornotemark[4]

\author{Weiqin Yang}
    \orcid{0000-0002-5750-5515}
    \affiliation{
        \institution{Zhejiang University}
        \city{Hangzhou}
        \country{China}
    }
    \email{tinysnow@zju.edu.cn}
\authornotemark[1]
\authornotemark[3]
\authornotemark[4]

\author{Jiawei Chen}
    \orcid{0000-0002-4752-2629}
    \affiliation{
        \institution{Zhejiang University}
        \city{Hangzhou}
        \country{China}
    }
    \email{sleepyhunt@zju.edu.cn}
\authornote{Corresponding author.}
\authornote{State Key Laboratory of Blockchain and Data Security, Zhejiang University.}
\authornote{College of Computer Science and Technology, Zhejiang University.}
\authornote{Hangzhou High-Tech Zone (Binjiang) Institute of Blockchain and Data Security.}

\author{Peng Wu}
    \orcid{0000-0001-7154-8880}
    \affiliation{
        \institution{Beijing Technology and Business University}
        \city{Beijing}
        \country{China}
    }
    \email{pengwu@btbu.edu.cn}
\authornote{School of Mathematics and Statistics, Beijing Technology and Business University.}

\author{Yuegang Sun}
    \orcid{0009-0009-2701-4641}
    \affiliation{
        \institution{Intelligence Indeed}
        \city{Hangzhou}
        \country{China}
    }
    \email{bulutuo@i-i.ai}

\author{Gang Wang}
    \orcid{0000-0001-6248-1426}
    \affiliation{
        \institution{Bangsheng Technology Co,Ltd.}
        \city{Hangzhou}
        \country{China}
    }
    \email{wanggang@bsfit.com.cn}

 \author{Qihao Shi}
     \orcid{0000-0002-7883-9848}
    \affiliation{
        \institution{Hangzhou City University}
        \city{Hangzhou}
        \country{China}
    }
    \email{shiqihao321@zju.edu.cn}

\author{Can Wang}
    \orcid{0000-0002-5890-4307}
    \affiliation{
        \institution{Zhejiang University}
        \city{Hangzhou}
        \country{China}
    }
    \email{wcan@zju.edu.cn}
\authornotemark[3]
\authornotemark[5]

\begin{abstract}
Recommender systems (RS) aim to retrieve a small set of items that best match individual user preferences. Naturally, RS place primary emphasis on the quality of the \topk results rather than performance across the entire item set. However, estimating \topk accuracy (\eg Precision@$K$, Recall@$K$) requires determining the ranking positions of items, which imposes substantial computational overhead and poses significant challenges for optimization. In addition, RS often suffer from distribution shifts due to evolving user preferences or data biases, further complicating the task.

To address these issues, we propose \lossName, a loss function that is specifically designed to optimize the \topk recommendation accuracy. \lossName\ leverages a quantile technique that replaces the complex ranking-dependent operations into simpler comparisons between predicted scores and learned score thresholds. We further develop a sampling-based regression algorithm for efficient and accurate threshold estimation, and introduce a constraint term to maintain optimization stability by preventing score inflation. Additionally, we incorporate a tailored surrogate function to address discontinuity and enhance robustness against distribution shifts. Comprehensive theoretical analyzes and empirical experiments are conducted to demonstrate the effectiveness, efficiency, convergence, and distributional robustness of \lossName. The code is available at \url{https://github.com/cynthia-shengjia/WWW-2026-Talos}.

\end{abstract}


 
\maketitle

\section{Introduction}\label{sec:introduction}

Being able to provide personalized suggestions, Recommender Systems (RS)~\cite{ko2022survey, zhang2019deep}  are integral to several online service platforms. On these platforms, users are typically presented with only a small set of items. Consequently, RS primarily emphasizes the quality of the \topk items (\eg Precision@$K$ and Recall@$K$) rather than the performance across the entire item set. Existing RS often adopt a learning-based paradigm~\cite{he2020lightgcn,wu2024bsl} --- learning a recommendation model to estimate user preference scores over items and then retrieving the \topk items with the highest predicted scores as recommendations.

The distinct focus on \topk accuracy has inspired extensive research on recommendation loss functions. The loss function determines the direction of model optimization, whose importance cannot be overemphasized~\cite{wu2024effectiveness,yang2024psl}. Recent years have witnessed the emergence of two prominent types of loss functions:

\begin{figure*}[t]
\centering
\includegraphics[width=1\textwidth]{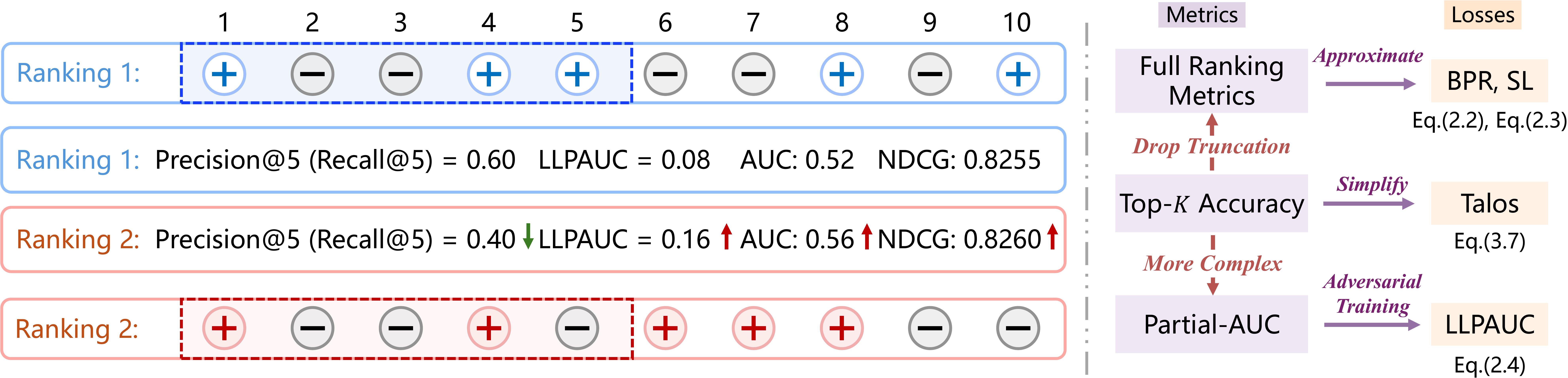}
    \caption{(\textit{Left}). Illustration of the inconsistency between LLPAUC/AUC/NDCG and Top-$K$ accuracy (Precision@$K$ and Recall@$K$) for two difference ranking schemes of ten items, where ranking scheme 2 achieves better AUC/LLPAUC and NDCG, while worse on \topk accuracy; (\textit{Right}). The relationship among \lossName, LLPAUC, SL, and BPR.}
    \label{fig-case-study}
\end{figure*}

\begin{itemize}[leftmargin = *]
\item \textbf{Full-ranking Losses}: Some works aim to improve the overall ranking performance without explicitly targeting \topk estimation. Prominent examples include BPR~\citep{rendle2009bpr} and Softmax Losses~\citep{wu2024effectiveness}, which have been shown to approximately optimize the Area Under the ROC Curve (AUC) and Normalized Discounted Cumulative Gain (NDCG), respectively. However, AUC and NDCG evaluate the quality of the entire ranking list, which can differ substantially from the quality of the \topk subset most relevant to RS outcomes. Consequently, optimizing these full-ranking loss functions does not necessarily translate to improvements in \topk accuracy and may sometimes reduce it. \Cref{fig-case-study} provides an example where AUC and NDCG increase while Precision@$K$ and Recall@$K$ decrease. Such cases are not rare in RS. By empirically analyzing arbitrary pairs of ranking lists for users on real-world datasets, we observe that the ratio of such inconsistent cases exceeds 34.37\% and 28.49\% on average.

\item \textbf{Partial-AUC-based Losses}: Another line of research focuses on developing loss functions for optimizing Partial-AUC metrics (\eg LLPAUC~\cite{shi2024lower} and OPAUC~\cite{shi2023theories}), which evaluate the partial area under the ROC curve and correlate more strongly with \topk accuracy than full‑ranking losses. Indeed, as shown in \Cref{fig:incosistency-between-topk-metrics}, the inconsistency ratio between LLPAUC and Precision@$K$ decreases to 21.86\% on average. Nevertheless, the gap still exists. More critically, Partial-AUC metrics are highly complex, incurring significant optimization challenges. It can be seen from their surrogate loss functions often involve adversarial training~\cite{shi2024lower}, which complicates training stability and reduces effectiveness. Furthermore, these loss functions introduce additional hyperparameters compared to BPR and SL, necessitating expensive and time-consuming tuning efforts. These limitations significantly hinder their practical application.
\end{itemize}

Given these limitations, an important research question raises: \textit{How can we design a loss function that directly optimizes top‑K accuracy in recommender systems?} Considering the characteristics of \topk optimization and RS, we identify the following key obstacles:
\begin{itemize}[leftmargin=*]
\item \textbf{Ranking-aware Discontinuous Objective}: \topk accuracy (\eg Precision@$K$) are computed only over the \topk ranked items, requiring a ranking‑dependent truncation --- determining which items lie in the \topk positions.  This process requires computing item ranking positions, incurring high computational cost. It also makes the objective discontinuous or constant in large regions, making gradient‑based optimization ineffective.

\item \textbf{Distributional-Shift Challenge:} It is important to note that RS often faces severe distribution shifts due to evolving user preference ~\cite{wang2022causal} or data collection biases~\cite{chen2020bias, chen2021autodebias, gao2023alleviating,lin2025recommendation}.  Loss functions that are robust to such shifts have been shown to be essential for RS performance~\cite{wu2024bsl}.
\end{itemize}

\begin{table*}[t]
\centering

\caption{Empirical analyses of the inconsistency between AUC/LLPAUC/NDCG/NDCG@$K$ and \topk accuracy across arbitrary pairs of ranking lists on real-world datasets.}
\label{fig:incosistency-between-topk-metrics}
\setlength{\tabcolsep}{0.8mm}
\begin{tabular}{l|cc|cc|cc|cc} 
\Xhline{1.3pt}
\multirow{2}{*}{\textbf{Metric}} & \multicolumn{2}{c|}{\textbf{Gowalla}} & \multicolumn{2}{c|}{\textbf{Beauty}}  & \multicolumn{2}{c|}{\textbf{Games}}  & \multicolumn{2}{c}{\textbf{Electronics}}   \\ 
\cline{2-9}
                      & \textbf{Precision@20} & \textbf{Recall@20}                & \textbf{Precision@20} & \textbf{Recall@20}                                     & \textbf{Precision@20} & \textbf{Recall@20}                                      & \textbf{Precision@20} & \textbf{Recall@20}                \\ 
\hline
AUC        & 33.17\%    & 33.89\%     & 34.67\%    & 35.46\%     & 35.17\%    & 34.19\%      & 33.32\%    & 34.58\%          \bigstrut[t]\\
LLPAUC     & 18.53\%    & 18.65\%     & 23.81\%    & 22.95\%     & 23.30\%    & 22.34\%      &23.30\%     & 22.25\%                      \\
NDCG       & 26.70\%    & 26.88\%     & 29.65\%    & 29.52\%     & 29.52\%    & 28.02\%      & 29.11\%    & 28.53\%                       \\
NDCG@$K$   & 19.01\%    & 18.24\%     & 22.02\%    & 22.55\%     & 21.14\%    & 22.21\%      & 17.65\%    & 17.51\%        \bigstrut[b]\\
\Xhline{1.3pt}
\end{tabular}
\end{table*}

To address these challenges, we propose \lossName, a new loss function specifically designed for optimizing \topk recommendation accuracy (\eg Precision@$K$ and Recall@$K$). To overcome the ranking-dependence, we introduce the quantile ~\cite{koenker2005quantile,hao2007quantile,shao2008mathematical} to reformulate \topk selection into simple comparisons between item scores and a learned score threshold. Items with scores above the threshold constitute the Top-$K$, avoiding explicit sorting and thus significantly facilitating optimization. However, this threshold-based strategy also raises two further questions: 1) \textbf{Threshold estimation}: The valid thresholds vary per user, requiring efficient and accurate estimation; 2) \textbf{Convergence concern}: Thresholds evolve during training, raising concerns about training convergence and stability. Particularly, we observe score inflation phenomenon that both predicted scores and thresholds rise excessively. We resolve these by: 1) Developing a sampling‑based quantile regression method for fast and accurate estimation; 2) Introducing a constraint term limiting the number of items above the threshold, ensuring training convergence and stability.

Additionally, to further tackle discontinuity and distribution shift, we design a specific smooth surrogate function approximating the Heaviside step function. These designs endow \lossName\ with three fundamental theoretical properties: 1) \lossName~serves as a tight upper bound for optimizing Precision@$K$, ensuring its effectiveness; 2) \lossName~is equivalent to performing \emph{Distributionally Robust Optimization}, a theoretical-sound approach that empowers the model with robustness against distribution shifts; 3) the optimization procedure for \lossName\ provides theoretically convergence guarantees.

The proposed \lossName\  is practical in three aspects: 1) It is concise in form and can be easily plug‑in existing RS models with minimal code modifications; 2) It is computationally efficient, with both theoretical time complexity and practical runtime comparable to the conventional loss; 3) It preserves simplicity in hyperparameter tuning, requiring only a single temperature parameter. In summary, our main contributions are:

\begin{itemize}[leftmargin=*]
\item We propose \lossName, a novel loss function for directly optimizing \topk recommendation accuracy.

\item We provide comprehensive theoretical analyses, demonstrating its effectiveness, distributional robustness, and convergence guarantees.

\item  We conduct extensive experiments across four datasets and three backbone models, verifying the superiority of \lossName\ over the state‑of‑the‑art losses.
\end{itemize}

\section{Preliminaries}

In this section, we present the task formulation, an overview of the \topk accuracy and an analysis of existing loss functions.

\subsection{Task Formulation}

Aligned with recent studies~\cite{he2017ncf,shi2014collaborative} on recommendation losses, this work focuses on the conventional collaborative filtering (CF) scenario. Let $\userSet$ and $\itemSet$ denote the sets of users and items, respectively. The set of observed interactions is represented as $\obSet \subseteq \userSet \times \itemSet$, where each entry $(u,i) \in \obSet$ signifies an interaction (\eg click, purchase, \etc) between user $u$ and item $i$. Consequently, for a given user $u$, the positive item set $\posU = \{i \in \itemSet : (u,i) \in \obSet\}$ is constructed from the observed interactions, while the remaining unobserved items constitute the negative item set $\negU = \itemSet \setminus \posU$. The primary objective of recommender systems is to leverage the historical data $\obSet$ to predict the \topk items that best match the user's preferences.

Existing RS mainly adopt a learning-based paradigm~\cite{he2017ncf}. Specifically, a parameterized recommendation model $M_\theta$ takes the features (\eg IDs) of user $u$ and item $i$ as input~\citep{yang2025breaking} to generate the preference scores $\sui=M_\theta(u,i)$. The model is then optimized from $\obSet$ under a specific loss function. Ultimately, the recommendation list for user $u$ is generated by ranking items according to $\sui$ and retaining the \topk candidates. Thus, the loss function, which governs the optimization trajectory of the model, plays a crucial role in RS.

\subsection{\topk Accuracy in RS}

As RS typically display only the \topk items to users, \topk accuracy, \eg Precision@$K$ and Recall@$K$, are commonly used to evaluate the recommendation performance, which are defined as
\begin{equation}
\label{eqs:topk-metrics}
    \text{Precision@}K(u)  \!= \!\!\!\sumPosU\!\!\frac{\I(\pi_{ui}\le K)}{K},
    \
    \text{Recall@}K(u)  \!= \!\!\!\sumPosU\!\!\frac{\I(\pi_{ui} \le K)}{\vert \posU \vert} 
\end{equation}
where $\I(\cdot)$ denotes the indicator function, $\pi_{ui} = \sum_{j\in\itemU}\I(\suj\ge\sui)$ denotes the ranking position of item $i$ for user $u$. These metrics measure the the quality (positivity) of the top-ranked items, and incorporate a ranking-dependent truncation term $\I(\pi_{ui}\le K)$ that determines whether an item ranks within the \topk. Notably, this study primarily targets the optimization of \topk accuracy, \ie Precision@$K$ and Recall@$K$. Although more sophisticated metrics (\eg NDCG@$K$, MRR@$K$) additionally capture position‑aware relevance, they are not as prevalent as Precision@$K$ and Recall@$K$ in modern RS. Precision@$K$ and Recall@$K$ exhibit a more direct correlation with online business‑critical metrics~\cite{covington2016deep,zhou2018deep,pi2020search} (\eg CTR), and are widely regarded as gold-standard indicators for the recall stage of recommendation pipelines~\cite{zhang2021learning,ren2024information}. Furthermore, since complex metrics are on the basis of \topk accuracy, improving \topk accuracy can potentially enhance these metrics, as evidenced by the results reported in \Cref{appendix-on-MRR-and-NDCG}.

\subsection{Analyses on Existing Loss Functions}\label{sec:preliminary-sl@k-disadvantage}

Beyond the traditional point-wise and pair-wise loss functions~\cite{pan2008one,johnson2014logistic,rendle2009bpr}, recent years have witnessed the emergence of two
prominent types of loss function in RS:

\noindent\textbf{Full Ranking Losses.} The most prominent examples that optimize the full ranking metrics are BPR and Softmax Loss (SL). BPR is a surrogate loss that is defined as:
\begin{equation}\label{eqs:bpr and original AUC}
    \loss_{\text{BPR}} = \! \frac{1}{\vert\posU\vert}\!\!\sum\limits_{i\in\posU}\sum\limits_{j\in\negU}\!\!\log\sigma(\suj-\sui), \ \   \text{AUC} =   \!\!\sum\limits_{i\in\posU}\sum\limits_{j\in\negU}\!\!\frac{\mathbb{I}(\sui\ge\suj)}{\vert\posU\vert \vert\negU\vert}
\end{equation}
where $\sigma$ seeks to approximate the Heaviside function, typically set as the sigmoid function~\citep{rendle2009bpr}. Consequently, BPR loss serves as an approximation of AUC metric. For each user, SL~\cite{wu2024effectiveness} is defined as:
\begin{equation}
\loss_{\text{SL}} = -\frac{1}{\vert \posU \vert}\sumPosU \log\frac{\exp(\sui/\tau)}{\sum_{j\in\itemU}\exp(\suj/\tau)}    
\end{equation}
where $\tau$ serves as a temperature hyperparameter. Recent theoretical analysis~\cite{yang2024psl} have demonstrated that SL as a tight surrogate loss for optimizing the full-ranking metric NDCG. This property often allows SL to yield state-of-the-art performance.

However, we argue that the full-ranking metrics differ significantly from \topk accuracy. This discrepancy is evident as \topk accuracy incorporate a truncation term $\I(\pi_{ui}\le K)$, indicating that performance is evaluated only on \topk items. To further illustrate this difference, we conduct an empirical analysis (see \Cref{appendix-simulation-detials}  for details). As shown in \Cref{fig:incosistency-between-topk-metrics}, the deviation between NDCG and \topk accuracy significantly  exceeds $20.47$\% on average. These results suggest that optimizing NDCG may not always lead to better \topk accuracy and, in some cases, may even degrade performance. As shown in \Cref{fig:incosistency-between-topk-metrics}, the deviation between NDCG and \topk accuracy is significant ($20.47$\% on average).

\noindent\textbf{Partial-AUC-based Losses.} 
Another research line focuses on  optimizing Partial-AUC metrics~\cite{zhu2022auc,shi2023theories}, which quantify the area under a specific segment of the ROC curve. The general Partial-AUC metric, Lower-Left Partial AUC (LLPAUC), is formally defined as~\cite{shi2024lower}:
\begin{equation}\label{eqs:LLPAUC-formula} 
    \begin{aligned}
    &\text{LLPAUC} = \sumPosU\sumNegU \frac{\I(\sui\ge\suj)\cdot\I(\sui\ge\eta_{\alpha})\cdot\I(\suj\ge\eta_{\beta})}{\vert \posU \vert \vert \negU \vert} \\ 
    &\text{\emph{s.t.}} \ \textbf{Pr}_{i\sim U(\posU)}[\sui \ge \eta_{\alpha}] = \alpha\ \ \text{and}\ \   \textbf{Pr}_{j\sim U(\negU)}[\suj \ge \eta_{\beta}] = \beta  
\end{aligned}
\end{equation}
where $\eta_{\alpha}$ and $\eta_{\beta}$ denote the hyperparameters that delineate the evaluation region, $U(\cdot)$ denotes the uniform distribution. Compared to the original AUC, constraints in \eq{\ref{eqs:LLPAUC-formula}} are introduced, evaluating ranking performance on positive and negative items with high scores. Recent work~\cite{shi2024lower} and our empirical analysis (\Cref{fig:incosistency-between-topk-metrics}) both show  Partial-AUC metrics are more consistent with \topk accuracy, with the average inconsistency ratio reduced to 21.86\%.

Nevertheless, we identify several critical limitations. First, the residual inconsistency between Partial-AUC and \topk accuracy remains non-negligible. Second, Partial-AUC metrics are highly complex, involving three indicator functions, significantly increasing the optimization difficulty. This complexity manifests in their surrogate loss functions, which often necessitate adversarial training~\cite{yang2024psl}, potentially compromising training stability and effectiveness. Our experimental observations confirm the instability of LLPAUC loss (\cf \Cref{table: overall performance}), which in some cases underperforms the basic BPR loss. Third, these losses introduce additional hyperparameters (\eg $\eta_{\alpha}$ and $\eta_{\beta}$ in \eqref{eqs:LLPAUC-formula}), requiring expensive tuning. These limitations heavily hinder the practical applications of these losses.

\noindent\textbf{NDCG@$K$ Surrogate Loss.} 
Recently, a contemporaneous and parallel study, SL@$K$~\cite{yang2025breaking}, has been proposed to optimize the \topk ranking metric NDCG@$K$. For each user, SL@$K$ is defined as:
\begin{equation} 
    \loss_{\text{SL@}K} = -\frac{1}{\vert\posU\vert}\sum\limits_{i\in\posU}\Big[\sigma\Big((\sui-\margin)/\tau_w\Big)\log\sum\limits_{j\in\itemSet}\exp\Big((\suj-\sui)/\tau\Big)\Big] 
\end{equation}
where $\tau_w$ and $\tau$ are two hyperparameters; $\margin$ is a user-specific score threshold (the score at rank $K$); $\sigma$ is the sigmoid function. SL@$K$ has been shown to serve as an upper bound for NDCG@$K$~\cite{yang2025breaking}.

However, several limitations warrant attention: 1) SL@$K$ is designed to optimize NDCG@$K$, rather than the \topk accuracy (\eg Precision@$K$ and Recall@$K$). These metrics differ substantially. As demonstrated in \Cref{fig:incosistency-between-topk-metrics}, NDCG@$K$ exhibits notable inconsistency with \topk accuracy (20.04\% on average).  Moreover, Precision@$K$ and Recall@$K$ are more
prevalent in industrial RS and align more closely with
common online business metrics~\cite{covington2016deep,zhou2018deep,pi2020search,cui2025field}. 2) SL@$K$ does not provide theoretical guarantees for robustness against distribution shifts, which is essential for RS~\cite{wu2024bsl}. 3) SL@$K$ demonstrates instability with fewer negative training signals (\cf \Cref{table: the varying negative numbers with SLatK}). 4) SL@$K$ relies on Monte-Carlo method to estimate the \topk threshold, which often incurs a high estimation error (0.18 on average, \cf \Cref{tab:error-of-quantile-estimation}). 5) SL@$K$ introduces an extra hyperparameter $\tau_w$ that requires exhaustive tuning. These observations motivate the development of loss functions tailored for \topk accuracy while addressing the aforementioned challenges.

\section{Methodology}
In this section, we first detail the proposed loss \lossName\ for optimizing \topk accuracy, and then we conduct comprehensive theoretical analyses to demonstrate its effectiveness.
\vspace{-5pt}
\subsection{Loss Function for \topk Accuracy}\label{methodology-of-tl-at-k}
Our \lossName\ first employs a quantile-based approach to simplify the ranking‑dependent truncation term. We subsequently design a regression method to enable efficient and accurate threshold estimation, and introduce an additional constraint term to ensure optimization stability. Finally, we integrate a customized surrogate function to mitigate discontinuities and improve distributional robustness.

% This section introduces the proposed loss for \topk metrics. For convenience, we mainly focus on Precision@$K$, as Recall@$K$ differs from Precision@$K$ only by a constant. Our experiments in \Cref{table: overall performance,table: overall performance on MRR and NDCG} also show that our loss improves Recall@$K$, NDCG@$K$ and MRR@$K$. Specifically, our \lossName~leverages the following strategies to tackle the \emph{truncation}, \emph{discontinuity}, and \emph{distributional-shift} challenges:

\noindent\textbf{Introducing Quantile Technique:} To tackle the non-feasible ranking-dependent truncation term, we borrow the quantile technique~\cite{koenker2005quantile,hao2007quantile,shao2008mathematical}. For each user, we introduce a score threshold named quantile that separates the \topk items from the rest according to their scores. The formal threshold can be defined as:
\begin{equation}
    \margin = \sui, \ \text{where}\ \  \pi_{ui} = K
\end{equation}
which represents the score of the item exactly at the $K$-th position.  When the item score exceeds the threshold, \ie $\sui\ge\margin$, it indicates that the item is included in Top-$K$, while $\sui\!\!<\!\!\margin$ implies it does not.

This technique simplifies the complex truncation term $\I(\pi_{ui} \!\le\! K)$ into a simple comparison between item scores and the threshold $\I(\sui \!\ge\! \margin)$, heavily facilitating fast computation and optimization. Specifically, the Precision@$K$\footnote{Considering Precision@$K$ may achieve the value 0, making the express $-\log{\text{Precision@}K}$ ill-defined. For rigor, we extend the support of the function and re-define $\log(x)$ as $\log(0)=\log(\epsilon)$, where $\epsilon$ denotes a sufficient small constant.}optimization can be transformed as:  
\begin{equation}\label{eqs:pre-at-k-original}
        -\log\text{Precision@}K = -\log\sumPosU\frac{\I(\sui\ge\margin)}{K}\\
\end{equation}
Here the objective only involves the model-predicted scores and thresholds. However, this quantile‑based transformation further raises two challenges: 1) \textbf{Threshold estimation}: the \topk threshold plays a central role in the optimization, and its precise estimation is crucial. Furthermore, the threshold must be computed individually for each user and evolves throughout training, thereby imposing significant computational demands. Conventional regression based strategies~\cite{koenker2005quantile} are computationally prohibitive in RS, while Monte Carlo‑based approaches~\cite{yang2025breaking} suffer from considerable estimation bias (\cf \Cref{tab:error-of-quantile-estimation}). Consequently, efficient and accurate threshold estimation in RS remains largely underexplored. 2) \textbf{Convergence concern}: The evolving threshold also poses risks to optimization convergence. In particular, we observe that directly optimizing \eq{\ref{eqs:pre-at-k-original}} often leads to pathological and unstable results, in which both positive and negative item scores, together with the thresholds, iteratively increase in a synchronized manner. This phenomenon, termed score inflation, is not desirable, as RS aim to differentiate positive and negative items and target positioning positive items at the Top-$K$. How to ensure the stable optimization is also an important question to be addressed.

% \Cref{appendix:quantile-regression} Appendix I.1
%  requires traversing the complete item space (\cf Appendix H.1),
%  \loss_{\text{QR}}(u)$, \cf Appendix H.1 for proof). 
\noindent\textbf{Efficient Quantile Estimation:} Quantile estimation~\citep{koenker2005quantile} has been well-studied in statistical learning literature. A prominent method is quantile regression~\citep{hao2007quantile}. However, estimating \topk quantile for a user in vanilla quantile regression requires traversing the complete item space (\cf \Cref{appendix:quantile-regression}), which is computationally intensive in large-scale RS. To address this, we propose an efficient negative sampling-based quantile regression loss:
\begin{equation} \label{eqs:debias-quantile-loss}
    \mathcal{Q}_K(u) = \frac{1}{\vert \itemSet \vert} \Big(\sum_{i\in\posU} \rho_{K}(\sui-\hat\beta_u)
    + w_u\sum_{j\in \sampledNegative} \rho_K(\suj-\hat\beta_u)\Big)  \\
\end{equation}
where $w_u \!=\! \frac{\vert\itemSet\vert - \vert\posU\vert}{\vert \sampledNegative \vert}$, $\rho_K(x) = (1 - K/\vert\itemSet\vert)(x)_+ +  (K/\vert\itemSet\vert) (-x)_+$ with $(x)_+ = \max\{0,x\}$, and $\sampledNegative$ denotes the sampled negative item set from $\negU$. We introduce importance weights $w_u$ to ensure unbiased estimation (\ie  $\mathbb{E}_{\sampledNegative}[\mathcal{Q}_K(u)] = \loss_{\text{QR}}(u)$, \cf \Cref{appendix:quantile-regression} for proof). This treatment can not only reduce the computational complexity to $\mathcal{O}(\vert\sampledNegative\vert + \vert \posU \vert)$ for a user, but remain high estimation accuracy. The estimation error is less than $0.02$ as reported in \Cref{tab:error-of-quantile-estimation}.

% thereby reducing the computational complexity to $\mathcal{O}(\vert\hat{\mathcal{N}}_u\vert + \vert \posU \vert)$ for a user. 

% where \ding{172} is due to the fact $\sum_{j\in\itemU}\I(\suj\ge\margin) = K$. Here $\delta(x)$ denotes the Heaviside step function with $\delta(x) = \I(x\ge0)$. The motivation behind this transformation can be explained as: Directly optimizing \eq{\ref{eqs:pre-at-k-original}} would easily sink into a degenerate and unstable solution, where scores $s_{ui}$ for positive and negative items are synchronously increased. This is not desirable, as RS aim to differentiate positive and negative items and target positioning positive items at the Top-$K$. The introduction of new denominator could naturally penalize the number of items larger than the \topk quantile, avoiding the score inflation phenomenon. From a theoretical view, equation \eq{\ref{eqs:pre-at-k-optimization}} can be understood as an objective equipped with constraint $\log\sum_{j\in\itemSet}\delta(\suj-\margin) = \log K$. Note that the Lagrange multiplier function of objective \eq{\ref{eqs:pre-at-k-original}} with this constraint can be written as follows:

\noindent\textbf{Constrained Optimization for Stable Convergence:} To tackle the inflation issue, we further transform \eq{\ref{eqs:pre-at-k-original}} into:
\begin{equation}\label{eqs:pre-at-k-optimization}
        \text{\eq{\ref{eqs:pre-at-k-original}}} = -\log\frac{\sum_{i\in\posU}\delta(\sui-\margin)}{\sum_{j\in\itemU}\delta(\suj-\margin)} 
\end{equation}
The equality holds due to the fact $\sum_{j\in\itemU}\I(\suj\ge\margin) = K$. Here $\delta(x)$ denotes the Heaviside step function with $\delta(x) = \I(x\ge0)$. The introduction of new denominator could naturally penalize the number of items larger than the \topk quantile, avoiding the score inflation phenomenon. From a theoretical view, \eq{\ref{eqs:pre-at-k-optimization}} can be understood as an objective equipped with constraint $\log\sum_{j\in\itemSet}\delta(\suj-\margin) = \log K$. Note that the Lagrange multiplier function of objective \eq{\ref{eqs:pre-at-k-original}} with this constraint can be written as follows:
\begin{equation}
    -\log\frac{1}{K}\sumPosU\delta(\sui-\margin)+\lambda (\log\sum_{j\in \itemU}\delta(\suj-\margin) -\log K )
\end{equation}
If we simply set the parameter $\lambda = 1$ and drop the irrelevant constant, we can obtain \eq{\ref{eqs:pre-at-k-optimization}}. The introduction of the denominator term helps mitigate the inflation of the score $s_{ui}$, as it enforces a penalty such that only $K$ items can exceed the quantile threshold. This constraint is crucial, and our empirical results demonstrate significant performance improvements when incorporating this denominator term (\cf \Cref{exp: ablation study}).

% \begin{table}[t]
% \caption{Training efficiency comparison (s/epoch).}
% \label{tabs:training-efficiency-in-two-real-world-dataset}
% \centering

% \begin{tabular}{l|cccc}
% \Xhline{1.3pt}
% \multicolumn{1}{c|}{\textbf{Loss}} & \multicolumn{1}{c}{\textbf{Gowalla}} & \multicolumn{1}{c}{\textbf{Beauty}} & \multicolumn{1}{c}{\textbf{Games}} & \multicolumn{1}{c}{\textbf{Electronics}} \bigstrut\\
% \Xhline{1.0pt}
%  SL   &  4.79       &       0.42       &       0.31       &       7.01                  \bigstrut[t]\\
%  PSL    &  4.96       &       0.42       &       0.35       &       7.15                                    \\
%   LLPAUC    &  4.82       &       0.52       &       0.37       &       6.63                              \\
%  \lossName\    &  5.37       &       0.54       &       0.43       &       7.98                             \bigstrut[b]\\

% \Xhline{1.3pt}
% \end{tabular}

% \end{table}

\noindent\textbf{Introducing Customized Surrogate Function:} Note that the discontinuity mainly arises from the Heaviside function $\delta(x)$. The conventional solution is to develop a proper activation function to approximate it as $\sigma(\cdot) \approx \delta(\cdot)$. Previous works \cite{iliev2017approximation,iliev2015approximation,yang2025breaking} typically introduce a temperature $\tau$ within the sigmoid function $\sigma(\cdot) = \text{sigmoid}(\cdot/\tau)$ to approximate the Heaviside step function. In this work, 
we incorporate $\tau$ outside the sigmoid function, \ie $\sigma_{\tau}(\cdot) = \text{sigmoid}(\cdot)^{1/\tau}$
and define our \lossName\ as follows:
\begin{equation} \label{eqs:our-original-loss}
    \loss_{\text{\lossName}} = -\log\frac{\sum_{i\in\posU}\sigma_{\tau}(\sui-\margin)}{\sum_{j\in\itemSet}\sigma_{\tau}(\suj-\margin)}
\end{equation}
This design has multiple advantages: 1) It equips \lossName\ with distributional robustness. In \Cref{sec:analyse-on-our-loss}, we demonstrate that optimizing \lossName~is equivalent to performing distributional robustness optimization, ensuring the model is optimized under distribution perturbations. However, the function with inner temperature cannot enjoy such a merit. 2) This design ensures \lossName~serving as an tight upper bound for $-\log\text{Precision@}K$ (\cf \Cref{theorem:the-upper-bound-property}), ensuring optimizing \lossName~can improve \topk accuracy. 3) Empirical results presented in \Cref{exp: ablation study} also demonstrate that activation with an outside temperature performs much better than inside.

\noindent\textbf{Implementation Details.} The optimization process alternates between two steps: 1) updating model parameters through gradient descent using \eq{\ref{eqs:our-original-loss}}; 2) updating quantile estimates via gradient descent using \eq{\ref{eqs:debias-quantile-loss}}. This procedure iterates until convergence. Besides, to address scalability challenges with large item sets, a negative sampling strategy is adopted in the calculation of \lossName, similar to SL and LLPAUC, where the denominator term in \eq{\ref{eqs:our-original-loss}} is computed over a sampled subset of negative items.

Finally, note that \lossName\ involves retrieving all positive items for each user, which may incur implementation complexity and does not facilitate parallel computation. Therefore, in practice, we simply change the order of the summation  and logarithmic  operations as:
\begin{equation} \label{eqs:TL-outside-sum}
    \loss_{\text{\lossName}} = -\frac{1}{\vert \posU\vert}\sumPosU\log\frac{\sigma_{\tau}(\sui-\margin)}{\sum_{j\in\sampledNegative}\sigma_{\tau}(\suj-\margin)}
\end{equation}
This formulation greatly facilitates implementation. We do not need to sum over all positive items in the system, which also facilitates mini-batch updates --- we can sample a mini-batch of positive items for optimization. In fact, this transformation has a theoretical basis where \eq{\ref{eqs:pre-at-k-optimization}} can be upper bounded by \eq{\ref{eqs:TL-outside-sum}} due to the Jensen's inequality~\cite{jensen1906fonctions}.
% \begin{equation}
%     \text{\eq{\ref{eqs:pre-at-k-optimization}}} \le -\frac{1}{\vert\posU\vert}\sumPosU\log\frac{\delta(\sui-\margin)}{\sum_{j\in\itemSet}\delta(\suj-\margin)} \le \text{\eq{\ref{eqs:TL-outside-sum}}}
% \end{equation}
% where we apply relaxation with  to facilitate implementation. Empirically, we find that this incurs only a limited performance impact.

\subsection{Analyses on \lossName}\label{sec:analyse-on-our-loss}
Our proposed \lossName~offers several advantages:

\noindent\textbf{Advantage 1: Concise and efficient.} \lossName\ has a concise form (\cf \eq{\ref{eqs:TL-outside-sum}}). Compared to SL, it differs only in integrating the quantile $\margin$ and replacing the activation from $\exp(x)^{1/\tau}$ to $\text{sigmoid}(x)^{1/\tau}$. Such a simple revision allows \lossName~to approximately optimize \topk accuracy. Similar to SL, only hyperparameter $\tau$ is introduced.

% ================= Shengjia Revise History ==========================
% the differences just lie in 
% ---------------->
% it differs only in
% ================= Shengjia Revise History ==========================

% \Cref{appendix-computational-efficiency}
% Our experiments also confirm the computational efficiency of \lossName\ (\cf Appendix G).
Regarding the efficiency, the time complexity of \lossName~is calculated as $\mathcal{O}(\bar{P}\vert\userSet\vert\vert\sampledNegative\vert)$ for both quantile estimation and loss optimization, which is the same as SL. Here, $\bar{P}$ denotes the average number of positive items per user; $\sampledNegative$ denotes the sampled negative item set satisfying $\vert \sampledNegative\vert \ll \vert \itemSet\vert$. Our experiments also confirm the computational efficiency of \lossName\ (\cf  \Cref{appendix-computational-efficiency}).

 \noindent\textbf{Advantage 2: Tight surrogate for optimizing Precision@$K$.} We establish theoretical connections between \lossName~ and Precision@$K$:

\begin{table*}[t]
\tiny
\centering
\caption{Overall performance comparison of \lossName\ with other losses. \colorbox{lightblue}{\textcolor{customblue}{blue}} indicates the best result, and the runner-up is underlined. \textcolor{customred}{Imp.\%} indicates the relatively improvements of \lossName\ over the best baselines. "P@20" denotes the metric Precision@20, and "R@20" denotes Recall@20; The mark `*' suggests the improvement is statistically significant with $p<0.05$. }
\label{table: overall performance}
\resizebox{\textwidth}{!}{
\begin{tabular}{c|l|cc|cc|cc|cc} 
\Xhline{1.5pt}
\multirow{2}{*}{\textbf{Model}} & \multicolumn{1}{c|}{\multirow{2}{*}{\textbf{Loss}}} & \multicolumn{2}{c|}{\textbf{Gowalla}}                           & \multicolumn{2}{c|}{\textbf{Beauty}}                     & \multicolumn{2}{c|}{\textbf{Games}}                 & \multicolumn{2}{c}{\textbf{Electronics}}                      \\ 
\cline{3-10}
& \multicolumn{1}{c|}{}  & \multicolumn{1}{c}{\textbf{P@20}} & \multicolumn{1}{c|}{\textbf{R@20}}   & \multicolumn{1}{c}{\textbf{P@20}} & \multicolumn{1}{c|}{\textbf{R@20}}  & \multicolumn{1}{c}{\textbf{P@20}} & \multicolumn{1}{c|}{\textbf{R@20}}   & \multicolumn{1}{c}{\textbf{P@20}} & \multicolumn{1}{c}{\textbf{R@20}}  \\ 
\Xhline{1.0pt}
\multirow{11}{*}{MF}      
&BPR &$0.0438$&$0.1511$&$0.0146$&$0.1267$&$0.0142$&$0.1334$&$0.0058$&$0.0566$\bigstrut[t]\\
&AATP &$0.0327$&$0.1078$&$0.0120$&$0.1044$&$0.0142$&$0.1330$&$0.0028$&$0.0272$ \\
&RS@$K$ &$0.0397$&$0.1218$&$0.0094$&$0.0669$&$0.0104$&$0.0872$&$0.0020$&$0.0159$ \\
&SmoothI@$K$ &$0.0586$&$0.1882$&$0.0163$&$0.1352$&$0.0192$&$0.1778$&$0.0059$&$0.0565$ \\
&SL &$0.0625$&$0.2017$&$0.0172$&$0.1403$&$0.0208$&$0.1918$&$0.0064$&$0.0626$ \\
&BSL &$0.0625$&$0.2017$&$0.0171$&$0.1401$&\underline{$0.0210$}&$0.1934$&$0.0065$&$0.0630$ \\
&PSL &\underline{$0.0631$}&\underline{$0.2031$}&\underline{$0.0174$}&\underline{$0.1432$}&$0.0210$&$0.1935$&\underline{$0.0066$}&\underline{$0.0637$} \\
&AdvInfoNCE &$0.0623$&$0.2012$&$0.0171$&$0.1403$&$0.0210$&\underline{$0.1935$}&$0.0063$&$0.0616$ \\
&LLPAUC  &$0.0562$&$0.1847$&$0.0157$&$0.1336$&$0.0187$&$0.1748$&$0.0059$&$0.0583$ \\
&\lossName&\cellcolor{lightblue}\textcolor{customblue}{$\textbf{0.0642}$}&\cellcolor{lightblue}\textcolor{customblue}{$\textbf{0.2079}$}&\cellcolor{lightblue}\textcolor{customblue}{$\textbf{0.0179}$}&\cellcolor{lightblue}\textcolor{customblue}{$\textbf{0.1499}$}&\cellcolor{lightblue}\textcolor{customblue}{$\textbf{0.0213}$}&\cellcolor{lightblue}\textcolor{customblue}{$\textbf{0.1967}$}&\cellcolor{lightblue}\textcolor{customblue}{$\textbf{0.0067}$}&\cellcolor{lightblue}\textcolor{customblue}{$\textbf{0.0655}$}\bigstrut[b]\\
\hhline{~*{9}{!{\arrayrulecolor{black}}-}} % 黑色虚线
&\textcolor{customred}{\textbf{Imp.\%}}&$\cellcolor{pink!20}\textcolor{customred}{\textbf{+1.71\%}^*}$&$\cellcolor{pink!20}\textcolor{customred}{\textbf{+2.35\%}^*}$&$\cellcolor{pink!20}\textcolor{customred}{\textbf{+2.69\%}^*}$&$\cellcolor{pink!20}\textcolor{customred}{\textbf{+4.69\%}^*}$&$\cellcolor{pink!20}\textcolor{customred}{\textbf{+1.19\%}^*}$&$\cellcolor{pink!20}\textcolor{customred}{\textbf{+1.62\%}^*}$&$\cellcolor{pink!20}\textcolor{customred}{\textbf{+3.04\%}^*}$&$\cellcolor{pink!20}\textcolor{customred}{\textbf{+2.75\%}^*}$\bigstrut\\
\Xhline{1.0pt}
\multirow{11}{*}{LGCN}      
&BPR&$0.0527$&$0.1761$&$0.0159$&$0.1385$&$0.0192$&$0.1801$&$0.0050$&$0.0482$ \bigstrut[t]\\
&AATP&$0.0249$&$0.0754$&$0.0121$&$0.1067$&$0.0127$&$0.1175$&$0.0030$&$0.0289$ \\
&RS@$K$&$0.0536$&$0.1735$&$0.0151$&$0.1192$&$0.0168$&$0.1511$&$0.0038$&$0.0357$ \\
&SmoothI@$K$&$0.0590$&$0.1898$&$0.0169$&$0.1427$&$0.0196$&$0.1801$&$0.0065$&$0.0630$ \\
&SL&$0.0628$&$0.2025$&$0.0172$&\underline{$0.1433$}&\underline{$0.0211$}&\underline{$0.1942$}&$0.0065$&$0.0629$ \\
&BSL&$0.0628$&$0.2025$&$0.0172$&$0.1433$&$0.0210$&$0.1931$&$0.0064$&$0.0626$ \\
&PSL&\underline{$0.0634$}&\underline{$0.2042$}&\underline{$0.0173$}&$0.1425$&$0.0210$&$0.1925$&$0.0065$&$0.0627$ \\
&AdvInfoNCE&$0.0627$&$0.2028$&$0.0172$&$0.1433$&$0.0210$&$0.1936$&$0.0064$&$0.0626$ \\
&LLPAUC&$0.0516$&$0.1722$&$0.0170$&$0.1423$&$0.0202$&$0.1882$&\underline{$0.0066$}&\underline{$0.0642$} \\
&\lossName&\cellcolor{lightblue}\textcolor{customblue}{$\textbf{0.0642}$}&\cellcolor{lightblue}\textcolor{customblue}{$\textbf{0.2080}$}&\cellcolor{lightblue}\textcolor{customblue}{$\textbf{0.0178}$}&\cellcolor{lightblue}\textcolor{customblue}{$\textbf{0.1489}$}&\cellcolor{lightblue}\textcolor{customblue}{$\textbf{0.0212}$}&\cellcolor{lightblue}\textcolor{customblue}{$\textbf{0.1960}$}&\cellcolor{lightblue}\textcolor{customblue}{$\textbf{0.0068}$}&\cellcolor{lightblue}\textcolor{customblue}{$\textbf{0.0662}$}\bigstrut[b]\\
\hhline{~*{9}{!{\arrayrulecolor{black}}-}} % 黑色虚线
&\textcolor{customred}{\textbf{Imp.\%}}&$\cellcolor{pink!20}\textcolor{customred}{\textbf{+1.26\%}^*}$&$\cellcolor{pink!20}\textcolor{customred}{\textbf{+1.85\%}^*}$&$\cellcolor{pink!20}\textcolor{customred}{\textbf{+3.08\%}^*}$&$\cellcolor{pink!20}\textcolor{customred}{\textbf{+3.87\%}^*}$&$\cellcolor{pink!20}\textcolor{customred}{\textbf{+0.68\%}^*}$&$\cellcolor{pink!20}\textcolor{customred}{\textbf{+0.95\%}^*}$&$\cellcolor{pink!20}\textcolor{customred}{\textbf{+3.48\%}^*}$&$\cellcolor{pink!20}\textcolor{customred}{\textbf{+3.06\%}^*}$\bigstrut\\
\Xhline{1.0pt}
\multirow{11}{*}{XSimGCL}      
&BPR&$0.0584$&$0.1917$&$0.0170$&\underline{$0.1450$}&$0.0194$&$0.1810$&$0.0064$&$0.0625$\bigstrut[t]\\
&AATP&$0.0451$&$0.1515$&$0.0153$&$0.1264$&$0.0173$&$0.1619$&$0.0051$&$0.0501$ \\
&RS@$K$&$0.0515$&$0.1669$&$0.0151$&$0.1221$&$0.0174$&$0.1590$&$0.0040$&$0.0373$ \\
&SmoothI@$K$&$0.0532$&$0.1731$&$0.0077$&$0.0560$&$0.0144$&$0.1320$&$0.0040$&$0.0395$ \\
&SL&$0.0623$&$0.2010$&$0.0171$&$0.1418$&$0.0208$&$0.1915$&$0.0065$&$0.0627$ \\
&BSL&$0.0625$&$0.2013$&$0.0169$&$0.1413$&\underline{$0.0209$}&\underline{$0.1931$}&$0.0063$&$0.0619$ \\
&PSL&\underline{$0.0630$}&\underline{$0.2023$}&\underline{$0.0172$}&$0.1407$&$0.0207$&$0.1910$&\underline{$0.0066$}&\underline{$0.0636$} \\
&AdvInfoNCE&$0.0623$&$0.2010$&$0.0171$&$0.1415$&$0.0209$&$0.1923$&$0.0064$&$0.0622$ \\
&LLPAUC&$0.0612$&$0.1980$&$0.0171$&$0.1443$&$0.0207$&$0.1910$&$0.0064$&$0.0635$ \\
&\lossName&\cellcolor{lightblue}\textcolor{customblue}{$\textbf{0.0638}$}&\cellcolor{lightblue}\textcolor{customblue}{$\textbf{0.2057}$}&\cellcolor{lightblue}\textcolor{customblue}{$\textbf{0.0180}$}&\cellcolor{lightblue}\textcolor{customblue}{$\textbf{0.1506}$}&\cellcolor{lightblue}\textcolor{customblue}{$\textbf{0.0212}$}&\cellcolor{lightblue}\textcolor{customblue}{$\textbf{0.1961}$}&\cellcolor{lightblue}\textcolor{customblue}{$\textbf{0.0067}$}&\cellcolor{lightblue}\textcolor{customblue}{$\textbf{0.0654}$}\bigstrut[b]\\
\hhline{~*{9}{!{\arrayrulecolor{black}}-}} % 黑色虚线
&\textcolor{customred}{\textbf{Imp.\%}}&$\cellcolor{pink!20}\textcolor{customred}{\textbf{+1.27\%}^*}$&$\cellcolor{pink!20}\textcolor{customred}{\textbf{+1.68\%}^*}$&$\cellcolor{pink!20}\textcolor{customred}{\textbf{+4.87\%}^*}$&$\cellcolor{pink!20}\textcolor{customred}{\textbf{+3.81\%}^*}$&$\cellcolor{pink!20}\textcolor{customred}{\textbf{+1.32\%}^*}$&$\cellcolor{pink!20}\textcolor{customred}{\textbf{+1.60\%}^*}$&$\cellcolor{pink!20}\textcolor{customred}{\textbf{+2.65\%}^*}$&$\cellcolor{pink!20}\textcolor{customred}{\textbf{+2.81\%}^*}$\bigstrut\\
\Xhline{1.5pt}
\end{tabular}
}
\end{table*}

\begin{table*}[t]
    \centering
    \begin{minipage}[t]{0.48\textwidth}
    
\centering
        \captionof{table}{Performance comparisons with varying $K$ on MF backbone in Gowalla dataset. \colorbox{lightblue}{\textcolor{customblue}{blue}} indicates the best result.}
        \label{tab:varying-k-result-on-gowalla}
        \begin{tabular}{l|cccc}
            \Xhline{1.3pt}
            \multicolumn{1}{l|}{\textbf{Gowalla}}   & \textbf{Recall@20}  & \textbf{Recall@50} & \textbf{Recall@80} \bigstrut\\
            \hline
            AATP        &$0.1062$&$0.2068$&$0.2160$  \bigstrut[t] \\
            RS@$K$       &$0.1218$&$0.1059$&$0.2406$                       \\
            SmoothI@$K$  & $0.1882$ & $0.3027$ & $0.3625$                               \\
            SL                  &$0.2017$&$0.3185$&$0.3900$                             \\
            BSL                 &$0.2017$&$0.3185$&$0.3900$                             \\
            AdvInfoNCE          &$0.2013$&$0.3183$&$0.3900$                             \\
            LLPAUC              &$0.1847$&$0.2927$&$0.3640$                             \\
            PSL                 &$0.2038$&$0.3192$&$0.3902$                            \\
                \lossName@{20}         & \cellcolor{skyblue!50} \textcolor{customblue}{\textbf{0.2078}} & $0.3202$& $0.3894$     \\
                \lossName@{50}         & $0.2078$&  \cellcolor{skyblue!50} \textcolor{customblue}{\textbf{0.3212}} & $0.3907$    \\
                \lossName@{80}         & $0.2074$& $0.3211$&  \cellcolor{skyblue!50} \textcolor{customblue}{\textbf{0.3911}}    \bigstrut[b]\\            
                \hline
            \textcolor{customred}{\textbf{Imp.\%}} & $ \textcolor{customred}{\textbf{+1.98\%}}$ &$\textcolor{customred}{\textbf{+0.63\%}}$&$\textcolor{customred}{\textbf{+0.22\%}}$ \bigstrut\\
            \Xhline{1.3pt}
        \end{tabular}
    \end{minipage}
    \hfill
    \begin{minipage}[t]{0.48\textwidth}
        \centering
        \captionof{table}{Performance comparisons with varying $K$ on MF backbone in Beauty dataset. \colorbox{lightblue}{\textcolor{customblue}{blue}} indicates the best result.}
        
        \label{tab:varying-k-result}
        \begin{tabular}{l|cccc}
            \Xhline{1.3pt}
            \multicolumn{1}{l|}{\textbf{Beauty}}   & \textbf{Recall@20}  & \textbf{Recall@50} & \textbf{Recall@80} \bigstrut\\
            \hline
            AATP         &$0.1077$&$0.1850$&$0.2371$        \bigstrut[t]\\
            RS@$K$      & 0.0669 & 0.0989 & 0.1223 \\
            SmoothI@$K$ & 0.1352 & 0.2041 & 0.2454  \\

            SL              & 0.1403 & 0.2163 & 0.2617                           \\
            BSL             & 0.1401 & 0.2159 & 0.2617                          \\
            AdvInfoNCE      & 0.1409 & 0.2160 & 0.2621                           \\
            LLPAUC          & 0.1336 & 0.2047 & 0.2443                          \\
            PSL             & 0.1432 & 0.2184 & 0.2625                            \\
            \lossName@{20}         & \cellcolor{skyblue!50} \textcolor{customblue}{\textbf{0.1499}} & 0.2205 & 0.2620      \\
            \lossName@{50}         & 0.1484 & \cellcolor{skyblue!50} \textcolor{customblue}{\textbf{0.2218}} & 0.2637    \\
            \lossName@{80}         & 0.1485 & 0.2213 & \cellcolor{skyblue!50} \textcolor{customblue}{\textbf{0.2674}}    \bigstrut[b]\\
            \hline
            \textcolor{customred}{\textbf{Imp.\%}} & \textcolor{customred}{\textbf{+1.85\%}} & \textcolor{customred}{\textbf{+4.69\%}} & \textcolor{customred}{\textbf{+1.59\%}} \bigstrut\\
            \Xhline{1.3pt}
        \end{tabular}
    \end{minipage}
\end{table*}

\begin{table*}[t]
\centering
\captionof{table}{Temporal shift explorations. P@20 denotes Precision@20, while R@20 denotes Recall@20.}
\label{table: OOD-performance-comparison}
    \begin{tabular}{l|c|cccccccccc}
        \Xhline{1.3pt}
        \textbf{Dataset}& \textbf{Metric}  &\textbf{AATP} &\textbf{RS@$K$} & \textbf{SmoothI@$K$} & \textbf{SL} & \textbf{BSL} & \textbf{PSL} & \textbf{AdvInfoNCE} & \textbf{LLPAUC} & \textbf{\lossName} &  \textcolor{customred}{\textbf{Imp.\%}} \bigstrut\\
        \Xhline{1.0pt}
        \multirow{2}{*}{\textbf{Gowalla}} & \textbf{P@20} & 0.0300 & 0.0404 & 0.0544 & 0.0544 & 0.0544 & 0.0542 & 0.0542 & 0.0567 & \textbf{0.0574} & \textcolor{customred}{\textbf{+1.24\%}}\bigstrut[t]\\
        & \textbf{R@20} & 0.0807 & 0.1103 & 0.1497   & 0.1497 & 0.1497 & 0.1492 &0.1499 & 0.1541 &  \textbf{0.1577} & \textcolor{customred}{\textbf{+2.37\%}} \bigstrut[b]\\
        \hline
        \multirow{2}{*}{\textbf{Beauty}} & \textbf{P@20} & 0.0061 & 0.0046 & 0.0076  &  0.0089 & 0.0089 &  0.0086 & 0.0089 & 0.0087 &\textbf{0.0094} & \textcolor{customred}{\textbf{+5.51\%}}\bigstrut[t]\\
        & \textbf{R@20} & 0.0461 & 0.0381  & 0.0589 & 0.0696 & 0.0697 & 0.0679 & 0.0699 & 0.0697 & \textbf{0.0738} & \textcolor{customred}{\textbf{+5.45\%}}\bigstrut[b]\\
        \Xhline{1.3pt}
    \end{tabular}
\end{table*}

\begin{restatable}[\lossName~serves as a tight surrogate for       Precisio\-n@$K$]{theorem}{RestatableThmSurrogate} \label{theorem:the-upper-bound-property} For a proper $\tau$, satisfying $\tau \in [\frac{\log\left((e^2+2)\text{sigmoid}(-2)/2\right)}{\log\epsilon},\frac{\log(1/2)}{\log\epsilon}]$, we have the following bound relations:
\begin{equation}
    -\log\text{Precision@}K \le \loss_{\text{\lossName}} + C
\end{equation}
where $C = \log (1+e^2/2)^{1/\tau}$ is a constant. 
\end{restatable}
\noindent The proof is given in \Cref{appendix:proof-of-upper-bound}. This theorem ensures that optimizing \lossName\ can lower the upper bound of $-\log\text{Precision@}K$, and thus improve the Precision@$K$ with theoretical guarantee. 

\noindent\textbf{Advantage 3: Robustness to distribution shifts.} We establish a connection between \lossName\ and the \textit{Distributionally Robustness Optimization} (DRO). Specifically, we have:

\begin{restatable}[Distributional robustness]{theorem}{RestatableThmDRO} \label{theorem:DRO} For user $u$ and positive item $i$, let $\hat{Q}$ be the uniform negative distribution over $\negU$. Given a robustness radius $\eta \!>\! 0$, consider the uncertainty set $\mathbb{Q}$ consisting of all perturbed distributions $Q$, which is constrained by the KL divergence, i.e., $D_{\text{KL}}(Q\Vert\hat{Q}) \le \eta$. Let $\sigma(x)$ be a sigmoid function, optimizing \lossName\ is equivalent to solving the following optimization problem:
\begin{equation} \label{eq:TL-dro-equivalent}
     -\frac{1}{\vert \posU\vert}\sumPosU\log\sigma(\sui-\margin) + \max\limits_{Q\in\mathbb{Q}}\mathbb{E}_{j\sim Q}\Big[\log\sigma(\suj-\margin)\Big]  
\end{equation}
\end{restatable}
\noindent The proof is presented in \Cref{appendix:proof-of-distributional-robustness}. \Cref{theorem:DRO} demonstrates \lossName\ is equivalent to performing DRO~\cite{wu2023understanding,wu2024bsl,zhang2025advancing}, a theoretically sound tool that empowers models with robustness to distribution shifts. Intuitively, it can be understood that the model is trained not only on the original uniform negative distribution $\hat{Q}$, but across a family of adversarially perturbed distributions. Thus, the model trained under DRO would exhibit robustness. With proper design, our \lossName\ can inherit this property from DRO through equivalence. 

\noindent\textbf{Advantage 4: Convergence Guarantee.} While the proposed optimization process involves dynamic update between \eq{\ref{eqs:debias-quantile-loss}} and \eq{\ref{eqs:our-original-loss}}, it has the following theoretical convergence guarantee:

\begin{restatable}[Convergence guarantee]{theorem}{RestatableThmConvergence}
\label{theorem:convergence}
Let $K$ be the Lipschitz constant of $\loss_{\lossName}$. With the fixed stepsize $0 < \alpha < 2/K$, the gradient norm is bounded by:
\begin{equation}
\frac{(2\alpha - K\alpha^2)}{4}\mathbb{E}_t\left[\Vert \nabla \loss_{\text{\lossName}}^{(t)} \Vert^2  \right]\le   \frac{1}{T}\loss_{\lossName}^{(0)}
\end{equation}
where $T$ denotes the total number of optimization iterations, $\nabla \loss_{\text{\lossName}}^{(t)}$ denotes the gradient of the loss \text{w.r.t.} the score variables at iteration $t$, and $\loss_{\lossName}^{(0)}$ denotes the the initial loss value.
\end{restatable}
\noindent The proof is presented in \Cref{appendix:proof-of-convergence-property}. \Cref{theorem:convergence} 
provides a theoretical foundation for \lossName\ optimization reliability.

% \vspace{-11pt}
\section{Experiments}

\subsection{Experimental Setup}

 % \Cref{appendix-recommendation-backbone}
 % See Appendix E.2 for more details.
\textbf{Recommendation Backbones.} We closely refer to~\citet{yang2024psl} and evaluate \lossName\ and baselines  on three distinct recommendation backbones: the classic Matrix Factorization (MF) model~\citep{rendle2009bpr}, the representative graph-based model LightGCN~\citep{he2020lightgcn}, and the SOTA method XSimGCL~\citep{yu2023xsimgcl}. See \Cref{appendix-recommendation-backbone} for more details.

\noindent\textbf{Baselines.} The following baselines are included: 1) \textbf{BPR}~\cite{rendle2009bpr}, the classic pair-wise loss in RS; 2) \textbf{SL}~\cite{wu2024effectiveness}: the loss approximately optimizing the NDCG; 3) \textbf{BSL}~\cite{wu2024bsl}, \textbf{AdvInfoNCE}~\cite{zhang2024empowering}, and \textbf{PSL}~\cite{yang2024psl}: three enhanced variants of SL from different perspectives, achieving SOTA performance; 4) \textbf{LLPAUC}~\cite{shi2024lower}, the loss approximately optimizing the lower-left partial AUC; 5) \textbf{SmoothI@$K$}~\citep{SmoothIAAAI2021}, \textbf{RS@$K$}~\cite{patel2022recall}, and \textbf{AATP}~\cite{boyd2012accuracy}, three surrogate losses for \topk metrics. Note that these methods are tailored for other domains, and their effectiveness in RS is limited. We discuss these methods in \Cref{sec:related-work}.

% \Cref{appendix-dataset}
\noindent\textbf{Datasets.} We conduct experiments on four widely-used datasets containing users' ratings on items: \textbf{Beauty}, \textbf{Games},  \textbf{Electronics}~\citep{yang2024psl} and \textbf{Gowalla}~\cite{he2020lightgcn}. See  \Cref{appendix-dataset} for more details.

% \Cref{appendix-hyperparameter-setting} 
% We refer readers to Appendix D.5 for more details.
\noindent\textbf{Hyperparameter Setting.} A grid search is adopted to find optimal hyperparameters. For all compared methods, we closely follow configurations in their respective publications to ensure the optimal performance. We refer readers to \Cref{appendix-hyperparameter-setting} for more details.

 % \Cref{appendix-on-MRR-and-NDCG}
 \vspace{-5pt}
\subsection{Analysis on Experiment Details}
\textbf{\lossName~ VS. others.} \Cref{table: overall performance} compares \lossName\ with other baselines. Overall, \lossName\ outperforms all baselines across all datasets and backbones. Especially in Beauty, \lossName\ achieves the impressive improvements --- 3.83\% average on two evaluation metrics, highlighting the effectiveness of \lossName\ that directly optimizes \topk accuracy. Additional results in \Cref{appendix-on-MRR-and-NDCG} demonstrate \lossName\ also brings improvements on NDCG@$K$ and MRR@$K$.

\noindent\textbf{LLPAUC VS. SL.} While LLPAUC demonstrates a stronger correlation with \topk metrics than SL, we observe it does not consistently outperforms SL. In fact, LLPAUC is unstable across different backbones, even worse than basic BPR on Gowalla with LGCN.

\noindent\textbf{Performance comparison with varying $K$.} \Cref{tab:varying-k-result,tab:varying-k-result-on-gowalla} illustrate the performance across different values of $K$. We observe that \lossName\ consistently outperforms the compared methods for various $K$ values. However, as $K$ increases, the magnitude of the improvements decreases. This observation aligns with our intuition. As $K$ increases, the \topk accuracy gradually degrades to the full ranking metrics. Consequently, the advantage of optimizing for \topk accuracy diminishes as $K$ grows.

\noindent\textbf{Consistency study.} \Cref{tab:varying-k-result,tab:varying-k-result-on-gowalla} present Recall@$K'$, and the performance of \lossName@$K$ for varying values $K,K' \!\!\in\!\!\{20,50,80\}$. We observe that the best performance is achieved when $K\!=\!K'$, aligned with our expectations. For instance, when evaluating Recall@$80$ for a model trained with \lossName@$20$ (targets optimizing for Recall@$20$), the discrepancy between $K$ and $K'$ leads to performance drop.

\begin{figure*}[h]
    \caption{Sensitivity analysis of \lossName\ on $\tau$, where \colorbox{lightblue}{\textcolor{customblue}{-\ -\ -}} denotes the performance of SL with optimal hyperparameter.
    }
    \begin{subfigure}[b]{0.22\textwidth}
        \centering
        \includegraphics[width=\textwidth]{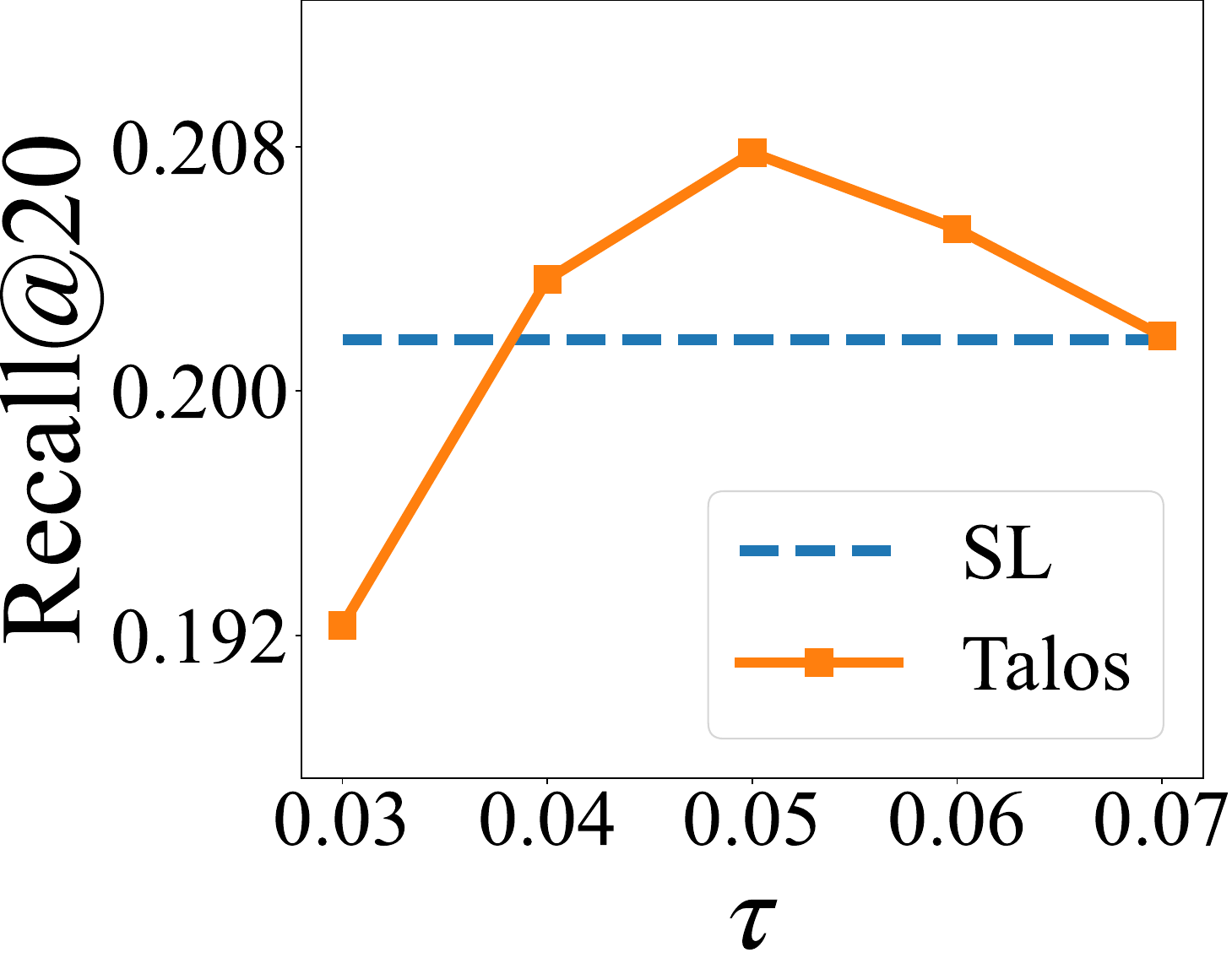}
        \caption{\hspace{0.3cm}Gowalla}
    \end{subfigure}
        \hfill
    \begin{subfigure}[b]{0.22\textwidth}
        \centering
        \includegraphics[width=\textwidth]{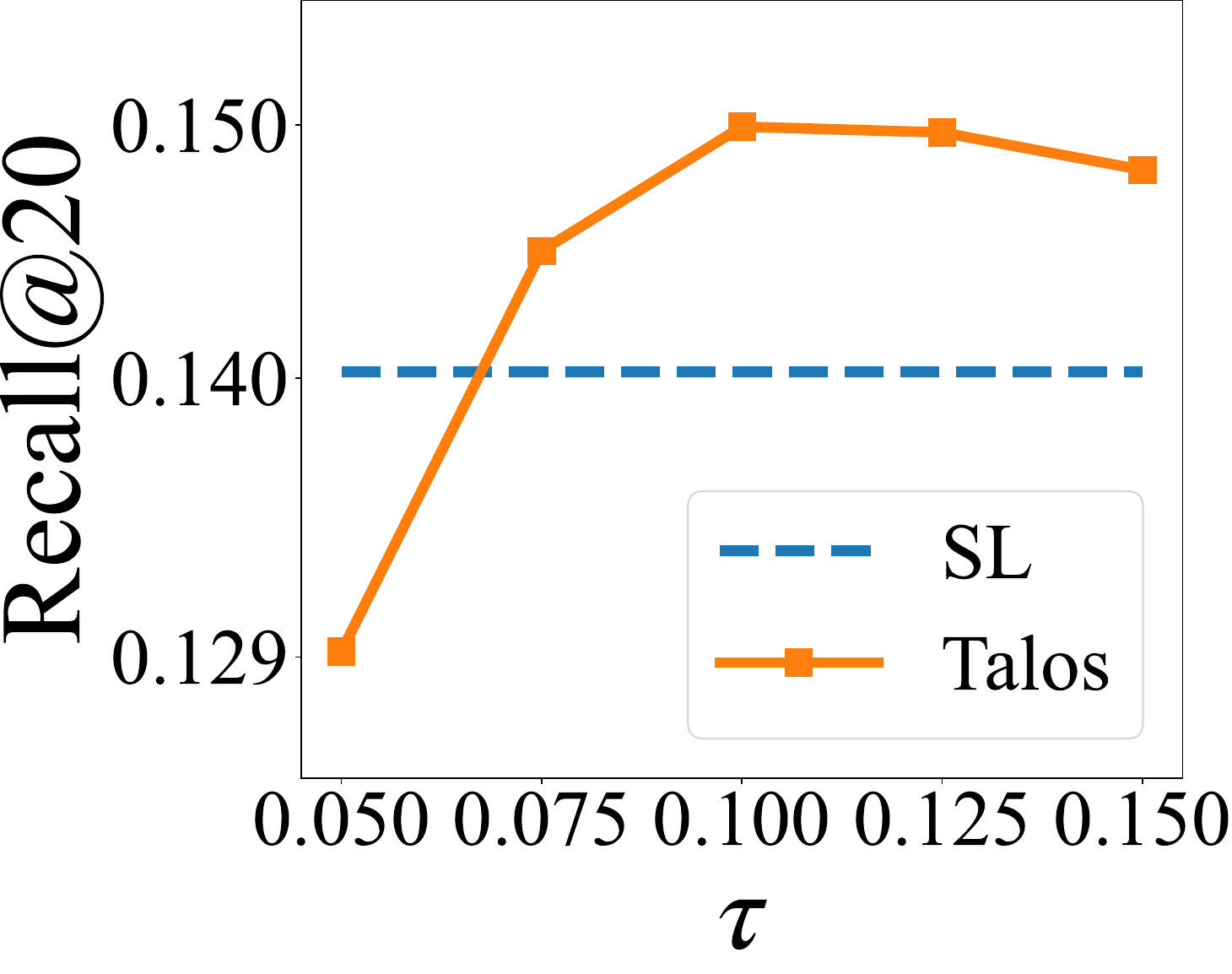}
        \caption{\hspace{0.3cm}Beauty}
    \end{subfigure}
    \hfill
    \begin{subfigure}[b]{0.22\textwidth}
        \centering
        \includegraphics[width=\textwidth]{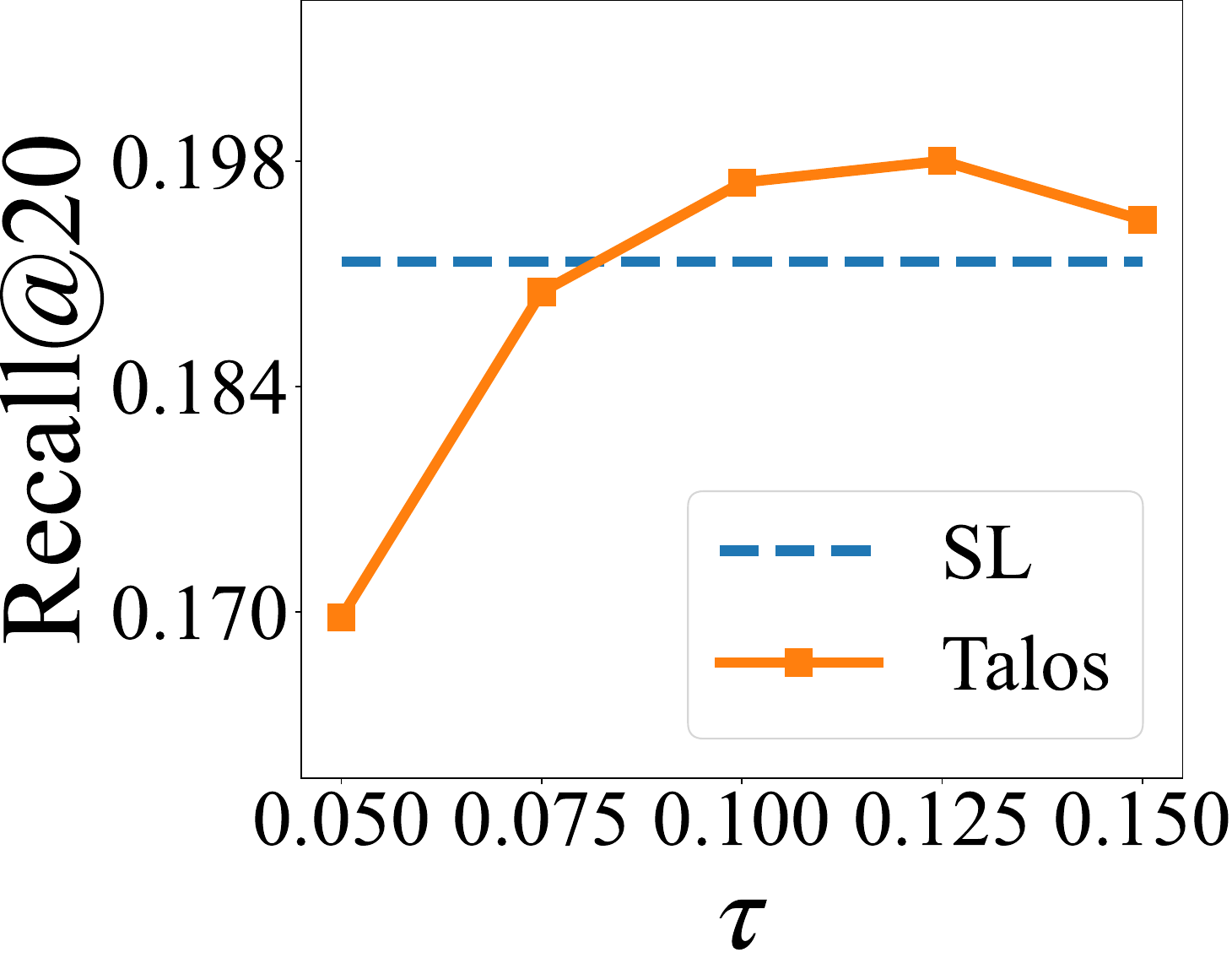}
        \caption{\hspace{0.3cm}Games}
    \end{subfigure}
    \hfill
    \begin{subfigure}[b]{0.22\textwidth}
        \centering
        \includegraphics[width=\textwidth]{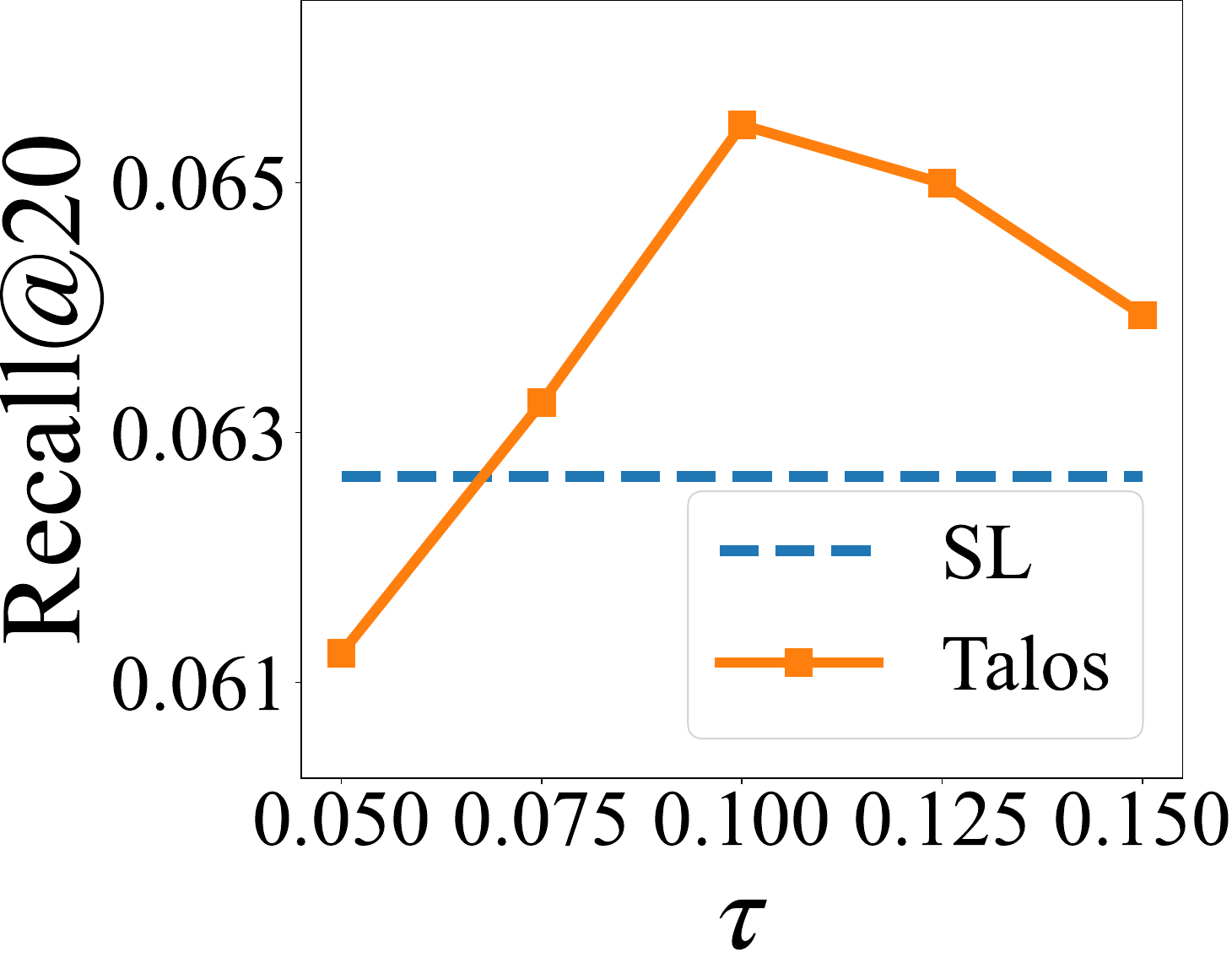}
        \caption{\hspace{0.2cm}Electronics}
    \end{subfigure}
    \label{appendix-temperature=sensitivity}
\end{figure*}
\begin{table*}[t]
% \scriptsize
\centering
\caption{Ablation study, we examine three variations of \lossName, where the quantile is removed (w/o-quantile), the $\text{sigmoid}(x)^{1/\tau}$ form is replaced by $\text{sigmoid}(x/\tau)$ (w/o-outside), and the denominator term is replaced by the constant $K$ (w/o-denominator).}
\label{exp: ablation study}
\resizebox{\textwidth}{!}{
\begin{tabular}{l|cc|cc|cc|cc} 
\Xhline{1.5pt}
\multirow{2}{*}{\textbf{Loss}} & \multicolumn{2}{c|}{\textbf{Gowalla}} & \multicolumn{2}{c|}{\textbf{Beauty}} & \multicolumn{2}{c|}{\textbf{Games}}  & \multicolumn{2}{c}{\textbf{Electronics}} \\ 
\cline{2-9}
& \textbf{Precision@20}   &    \textbf{Recall@20}       & \textbf{Precision@20}   &    \textbf{Recall@20}     &    \textbf{Precision@20}   &    \textbf{Recall@20}       &       \textbf{Precision@20}   &    \textbf{Recall@20}         \bigstrut\\ 
\Xhline{1.2pt}
 w/o-quantile   & 0.0626           & 0.2037             &   0.0175          &0.1487 & 0.0208            &0.1931&       0.0066      &0.0640 \bigstrut[t]\\
 w/o-outside    & 0.0577 
&  0.1859 
	  &   0.0166 
   &   0.1375 
 &   0.0194 
 &  0.1799 
 &0.0062& 	0.0592 

   \\
 w/o-denominator  & 0.0278           &0.0869 
              & 0.0101 
              & 0.0841 
    & 0.0127 
               & 0.1173 
  & 0.0018 
             & 0.0160 
   \\
 \lossName &$\textbf{0.0642}$  & $ \textbf{0.2079}$  & $\textbf{0.0179}$  & $\textbf{0.1499}$ &$\textbf{0.0213}$             &$\textbf{0.1967}$&$\textbf{0.0067}$             &$\textbf{0.0655}$  \bigstrut[b]\\
\Xhline{1.5pt}
\end{tabular}
}
\end{table*}

% \Cref{appendix-recommendation-scenarios}  test scenario with distribution shifts (\cf Appendix D.4).

\noindent\textbf{Performance with distribution shifts.} To evaluate the robustness against distribution shifts, we follow~\cite{wang2024distributionally}, introducing temporal bias to construct the test scenario with distribution shifts (\cf  \Cref{appendix-recommendation-scenarios}). \Cref{table: OOD-performance-comparison} shows that \lossName\ achieves the best performance, demonstrating its superior distributional robustness. LLPAUC benefits from adversarial training and exhibits fine robustness. Moreover, \lossName\ achieves more pronounced average improvements than in IID setting, \ie $1.26\%\!\to\! 3.37\%$. The effectiveness can be attributed to its connection with DRO.

\noindent\textbf{Ablation study.} In \Cref{exp: ablation study}, we evaluate three \lossName\ variants: 1) quantile fixed as zero (w/o-quantile); 2) $\text{sigmoid}(\cdot)^{1/\tau}$ replaced by $\text{sigmoid}(\cdot/\tau)$ (w/o-outside); 3) the denominator term in \eq{\ref{eqs:our-original-loss}} replaced by the constant $K$ (w/o-denominator). \lossName\ outperforms all variants. We highlight three key observations: 1) The gap between \lossName\ and w/o-quantile highlights the importance of quantile technique; 2) \lossName\ surpasses w/o-outside, indicating the superiority of introducing DRO; 3) Compared to w/o-denominator, results confirm our analysis in \Cref{methodology-of-tl-at-k} that \lossName\ naturally penalizes negatives exceeding the \topk quantile, addressing score inflation.

\noindent\textbf{Temperature Sensitivity.} \Cref{appendix-temperature=sensitivity} depicts the performance with varying $\tau$ in \lossName. Performance initially improves as $\tau$ increases, but degrades  beyond a certain point. This aligns with \Cref{theorem:the-upper-bound-property}: When $\tau$ satisfies the surrogate condition, smaller $\tau$ means the tighter upper bound for \topk accuracy but increases training difficulty due to the decreased Lipschitz smoothness. As $\tau$ increases, the approximation would be looser, even not satisfy the surrogate condition, thus impacting performance.

% In this section, we mainly introduce related work on loss functions. For recommendation models and loss consistency, please refer to \Cref{appendix-detailed-related-work}. In recommendation losses, two prominent types stand out: 1) full ranking optimization losses such as BPR~\citep{rendle2009bpr} and SL~\cite{wu2024effectiveness}. BPR optimizes the relative ranking of positive items over negatives as an AUC surrogate, while SL has been shown to approximately optimize the NDCG ranking metric with excellent performance. Motivated by the success of SL, several works attempted to improve it: BSL~\cite{wu2024bsl} and AdvInfoNCE~\cite{zhang2024empowering} introduced the DRO framework to enhance SL's robustness. PSL~\cite{yang2024psl} replaced the exponential function in SL with other appropriate activation functions, serving as a tighter surrogate for NDCG. 2) LLPAUC~\citep{shi2024lower}, which evaluates the lower left part of AUC, demonstrates a strong correlation with \topk accuracy. Beyond these two types, recent works explored alternative losses~\citep{sheng2023joint, lin2024understanding}, but these do not optimize \topk accuracy.

\section{Related Work}\label{sec:related-work}
Since this work focuses on recommendation loss functions, here we mainly introduce related work on this topic. For the area of recommendation models and recommendation loss consistency, we refer readers to \Cref{appendix-detailed-related-work}. In recommendation losses, two prominent types stand out: 1) full ranking optimization losses such as BPR~\citep{rendle2009bpr} and SL~\cite{wu2024effectiveness}. BPR optimizes the relative ranking of positive items over negatives as an AUC surrogate, while SL has been shown to approximately optimize the NDCG ranking metric with excellent performance. Motivated by the success of SL, several works attempted to improve it: BSL~\cite{wu2024bsl} and AdvInfoNCE~\cite{zhang2024empowering} introduced the DRO framework to enhance SL's robustness. PSL~\cite{yang2024psl} replaced the exponential function in SL with other appropriate activation functions, serving as a tighter surrogate for NDCG. 2) LLPAUC~\citep{shi2024lower}, which evaluates the lower left part of AUC, demonstrates a strong correlation with \topk accuracy.  Beyond these two types, recent works explored alternative losses~\citep{sheng2023joint, lin2024understanding}, but these do not optimize \topk accuracy.

% Beyond these two types, recent works have also explored other losses: \citet{sheng2023joint} employs two logits to optimize ranking; \citet{lin2024understanding} combines SL with BCE to resolve gradient vanishing issue. But these losses are not designed for optimizing \topk accuracy. 

Notably, \topk optimization has also been investigated in other fields, yet these approaches fall short when applied to RS. For instance, 
AATP~\cite{boyd2012accuracy} serves as a surrogate \topk loss by integrating quantile. Nevertheless, this loss is heuristically designed, which lacks theoretical connections to \topk accuracy. Furthermore, it fails to address the common distribution shifts in RS (\cf \Cref{table: OOD-performance-comparison}). In image retrieval, Pre@$K$~\cite{lu2019sampling} and RS@$K$~\cite{patel2022recall} optimize Precision@$K$ and Recall@$K$, respectively. However, they either integrated the complicated sampling strategy or rely on nested approximations of both ranking positions and indicator functions, which exacerbate errors when aligning with \topk accuracy. In information retrieval, SmoothI@$K$~\citep{SmoothIAAAI2021} employs the softmax function to approximate \topk indicators through a complex recursive estimation process, leading to inaccurate approximations particularly in RS with large item sets. As shown in \Cref{table: overall performance}, SmoothI@$K$ underperforms even the basic BPR when applied with the XSimGCL backbone.

Recently, a contemporaneous and parallel study, SL@$K$~\citep{yang2025breaking}, focuses on optimizing NDCG@$K$. We emphasize that Talos differs from SL@$K$ in several important aspects: 1) This work targets the optimization of Precision@$K$ and Recall@$K$, whereas SL@$K$ is designed to optimize NDCG@$K$. These metrics are not consistent as discussed in \Cref{sec:preliminary-sl@k-disadvantage}. 2) \lossName\ demonstrates the robustness to distribution shifts, while SL@$K$ does not. Under temporal shifts, \lossName\ achieves superior performance (over 1.12\%, \cf \Cref{table: the temporal shift scenarios with SLatK}). 3) \lossName\ achieves greater stability when fewer negative items are sampled, yielding markedly higher performance compared with SL@$K$ (over 1.96\%, \cf \Cref{table: the varying negative numbers with SLatK}). 4) \lossName\ incorporates a more accurate threshold estimation strategy, resulting in
substantially lower estimation error than SL@$K$ (0.01 v.s. 0.18, \cf \Cref{tab:error-of-quantile-estimation}). 5) SL@$K$ requires a greater number of hyperparameters, considerably limiting its practicality.

% including optimization targets, distributional robustness, stability with limited negative signals, quantile estimation precision, and hyperparameter complexity, as discussed in \Cref{sec:preliminary-sl@k-disadvantage}.

\section{Conclusion and Future Works}\label{sec:conclusions}

This work introduces \lossName, tailored for optimizing \topk accuracy in recommender systems. \lossName~incorporates an efficient quantile-based technique to handle the ranking-dependent challenge; introduces a constraint optimization term to address the score inflation issue; and leverage a specific surrogate function to tackle the discontinuity problem, equipping the loss with distributional robustness. Our theoretical analysis confirms the close bounded relationship between Precision@$K$ and \lossName, equivalence to distributionally robustness optimization, and convergence guarantees. Beyond these strengths, \lossName\ is concise in form and computationally efficient, making it practical in RS. Considering \lossName\ still introduces a temperature hyperparameter, developing adaptive $\tau$ mechanisms such as connecting $\tau$ to the number of positive interactions or using meta-learning, is a promising direction for future research.

\begin{acks}
This work is supported by the Zhejiang Province ``JianBingLingYan+X'' Research and Development Plan (2025C02020). We thank the reviewers for their valuable and insightful suggestions that improve the paper.
\end{acks}

\clearpage
\bibliographystyle{ACM-Reference-Format}
\bibliography{sample-base}

\appendix
% \setcounter{figure}{0}
% \setcounter{table}{0}
% \setcounter{equation}{0}

% \clearpage
\section{Theoretical Proofs}

\subsection{Proof of Theorem \ref{theorem:the-upper-bound-property}}\label[appendix]{appendix:proof-of-upper-bound}

% \RestatableThmSurrogate*

To prove \Cref{theorem:the-upper-bound-property}, we first derive \eq{\ref{eqs:pre-at-k-optimization}} as:
\begin{equation} \label{eqs:extend-topk-metric}
    -\log\!\text{Precision@}K \!=\!\colorbox{pink!25}{$\!\log\!\sum_{j\in\itemSet}\!\delta(\suj\!-\!\margin)$} - \colorbox{skyblue!25}{$\!\log\!\sum_{j\in\posU}\!\delta(\sui-\margin)$}  
\end{equation}
We simplify notation $\delta(\sui-\margin)$ and $\delta(\suj-\margin)$ as $\delta(x)$. Here, we simply consider $\sui \!\in\! [-1,\!1]$ and $x \!\!\in\!\! [-2,\!2]$ following recent work~\cite{wu2024bsl,yang2024psl}, as cosine similarity is commonly used to generate recommendation. Also, the normalization of prediction values has been demonstrated essential for RS~\cite{chen2023adap}. By selecting $\sigma(x) \!\!=\!\! \text{sigmoid}(x)^{1/\tau}$, for $\tau \!\!\ge\!\! \frac{\log\left((e^2+2)\text{sigmoid}(-2)/2\right)}{\log\epsilon} $, the \colorbox{pink!25}{\textcolor{customred}{red}} term can be bounded up as:
\begin{equation} \label{eqs:condition-one}
    \begin{cases}
    \log\delta(x) = \log(\epsilon) \le \log \sigma(x) + \log (1+e^2/2)^{1/\tau}  &, x \in [-2,0]\\
        \log\delta(x) \le \log(1)^{1/\tau}\le\log\sigma(x) + \log (1+e^2/2)^{1/\tau} &, x \in (0,2]          
    \end{cases}
\end{equation}
For $\tau$ satisfying $\tau \!\le\! \frac{\log(1/2)}{\log\epsilon}$, the \colorbox{skyblue!25}{\textcolor{customblue}{blue}} term can be bounded up as:
\begin{equation} \label{eqs:condition-two}
    \begin{cases}
        \log\sigma(x)  \le \log(1/2)^{1/\tau} \le \log\delta(x) & ,x \in [-2,0] \\
        \log\sigma(x)  \le \log(1) = \log\delta(x) & ,x \in (0,2]
    \end{cases}
\end{equation}
Thus, select a proper $\tau \in [\frac{\log\left((e^2+2)\text{sigmoid}(-2)/2\right) }{\log\epsilon},\frac{\log(1/2)}{\log\epsilon}]$, we have:
\begin{equation}
    -\log\text{Precision@}K \le \loss_{\text{\lossName}} + \log (1+e^2/2)^{1/\tau}
\end{equation}
\Cref{theorem:the-upper-bound-property} is proven. Notably, we observe the trade-off in temperature setting: 1) a large $\tau$ contradicting with $\tau \le \frac{\log(1/2)}{\log\epsilon}$ renders \eq{\ref{eqs:condition-two}} not hold, leading \lossName~ fails to serve as an upper bound; 2) $\tau$ is set too small would either contradict with \eq{\ref{eqs:condition-one}} or increase the constant $\log (1+e^2/2)^{1/\tau}$, scaling the gap with Precision@$K$.

\subsection{Proof of Theorem \ref{theorem:DRO}}
% \RestatableThmDRO*
\label[appendix]{appendix:proof-of-distributional-robustness} 

% The original DRO problem is defined as follows:
% \begin{equation} \label{eqs:Original-DRO-Problem}
%     \begin{aligned}
%      &-\frac{1}{\vert \posU\vert}\sumPosU\log\sigma(\sui-\margin) +\max\limits_{Q\in\mathbb{Q}}\mathbb{E}_{j\sim Q}\Big[\log\sigma(\suj-\margin)\Big] \\
%     \end{aligned}
% \end{equation}
% where $\mathbb{Q} = \{Q: D_{\text{KL}}(Q\Vert\hat{Q})\le\eta\}$. 

Given $-\frac{1}{\vert \posU\vert}\!\!\sum_{i\in\posU}\!\!\log\sigma(\sui-\margin)$ as constants do not affect optimization, the original DRO problem can be simplified  as:
\begin{equation}\label{eq:transformed-DRO}
 \max\limits_{Q\in\mathbb{Q}}\mathbb{E}_{j\sim Q}\Big[\log\sigma(\suj-\margin)\Big] \ \ \text{\emph{s.t.}}\ \  \mathbb{Q} = \{Q: D_{\text{KL}}(Q\Vert\hat{Q})\le\eta\}
\end{equation}
Let $L = Q(j)/\hat{Q}(j)$ and define a convex function $\phi(x) = x\log x - x + 1$, then the KL divergence can be written as $\mathbb{E}_{\hat{Q}}[\phi(L)]$. Let $g(u,j) = \log\sigma(\suj-\margin)$,  \eq{\ref{eq:transformed-DRO}} can be reformulated as:
\begin{equation}\label{eqs:Original-DRO-Problem}
        \max\limits_{L}\mathbb{E}_{\hat{Q}}[g(u,j)L] \ \  \text{\emph{s.t.}} \ \   \mathbb{E}_{\hat{Q}}[\phi(L)] \le \eta \ \ \text{and}\ \  \mathbb{E}_{\hat{Q}}[L] = 1
\end{equation}
As a convex optimization problem, we use the Lagrangian function to solve it:
\begin{equation} \label{eqs:DRO-intermediate}
    \!\!\!\min\limits_{\tau \ge 0, \lambda}\max\limits_{L}\left\{\mathbb{E}_{\hat{Q}}\Big[g(u,j)L\Big]  \!-\!\tau\left(\mathbb{E}_{\hat{Q}}\Big[\phi(L)\Big] \!-\! \eta\right) + \lambda\left(\mathbb{E}_{\hat{Q}}[L] \!-\! 1\right)\right\} 
\end{equation}
By the theorem of interchange of minimization and integration~\cite{rockafellar1998variational}, we can interchange maximization and expectation in \eq{\ref{eqs:DRO-intermediate}} as:
\begin{equation}\label{eqs:interexchange-max-and-sum}
        \text{\eq{\ref{eqs:DRO-intermediate}}} \iff \min\limits_{\tau\ge0, \lambda}\Big\{\tau\eta - \lambda + \tau\mathbb{E}_{\hat{Q}}\left[\max\limits_{L}\left\{\frac{g(u,j)+\lambda}{\tau}L - \phi(L)\right\}\right]\Big\}
\end{equation}
Notice that $\max\limits_{L}\left\{\frac{g(u,j)+\lambda}{\tau}L - \phi(L)\right\} = \phi^*(\frac{g(u,j)+\lambda}{\tau})$ is the convex conjugate function of $\phi(x)$, and we have $\phi^*(x) = e^x - 1$. Then:
\begin{equation} \label{eqs:DRO-finnal-subsititue}
        \text{\eq{\ref{eqs:interexchange-max-and-sum}}} = \min\limits_{\tau\ge0,\lambda}\left\{\tau\eta - \lambda + \tau \mathbb{E}_{\hat{Q}}\Big[\exp(\frac{g(u,j)+\lambda}{\tau}) - 1\Big]\right\} 
\end{equation}
The optimal $\lambda^*$, which minimizes the preceding expression is $\lambda^* = -\tau\log\mathbb{E}_{\hat{Q}}\Big[\exp(g(u,j)/\tau)\Big]$.
Ultimately, substituting $\lambda^*$, $g(u,j) =\log\sigma(\suj-\margin)$ back into \eq{\ref{eqs:DRO-finnal-subsititue}}, we have:
\begin{equation}
\begin{aligned}
        \text{\eq{\ref{eqs:DRO-finnal-subsititue}}} =& \min\limits_{\tau\ge0}\Big\{\tau\eta + \tau\log\mathbb{E}_{\hat{Q}}\left[\exp(\frac{g(u,j)}{\tau})\right]\Big\} \\
    =&  \min\limits_{\tau\ge0}\{\tau\eta + \tau\log\mathbb{E}_{\hat{Q}}\left[\sigma(\suj-\margin)^{1/\tau}\right]\}
\end{aligned}
\end{equation}
\eq{\ref{eq:TL-dro-equivalent}} is equivalent to optimizing:
\[
\begin{aligned}
    &-\frac{1}{\vert \posU\vert}\sumPosU\log\sigma(\sui-\margin) + \tau\log\mathbb{E}_{\hat{Q}}\left[\sigma(\suj-\margin)^{1/\tau}\right] \\
    &=  \tau\cdot \Big(\underbrace{-\frac{1}{\vert \posU\vert}\sumPosU\frac{1}{\tau}\log\sigma(\sui-\margin) + \log\mathbb{E}_{\hat{Q}}\left[\sigma(\suj-\margin)^{1/\tau}\right]}_{\text{\lossName}}\Big)
    % &=     \tau\cdot\underbrace{\left(-\frac{1}{\vert\posU\vert}\sumPosU\log\frac{\sigma(\sui-\margin)^{1/\tau}}{\E_{\hat{Q}}[\sigma(\suj-\margin)^{1/\tau}]}\right)}_{\loss_{\text{\lossName}}}
\end{aligned}
\]

\subsection{Proof of Theorem \ref{theorem:convergence}}\label[appendix]{appendix:proof-of-convergence-property}

We begin by calculating the accurate gradient \wrt $\mathbf{s}_u$ (the vector of all predicted scores) as follows:
\[
\nabla_{\mathbf{s}_u} \mathcal{L}_{\text{Talos}}(\mathbf{s}_u, \beta_u^K(\mathbf{s}_u))
= \frac{\partial \mathcal{L}_{\text{Talos}}}{\partial \mathbf{s}_u} + \frac{\partial \mathcal{L}_{\text{Talos}}}{\partial \beta_u^K} \frac{\partial \beta_u^K}{\partial \mathbf{s}_u}
\]
For simplicity, let $g = \frac{\partial \mathcal{L}_{\text{Talos}}}{\partial \mathbf{s}_u}$ and $q = \frac{\partial \mathcal{L}_{\text{Talos}}}{\partial \beta_u^K} \frac{\partial \beta_u^K}{\partial \mathbf{s}_u}$. In practical implementation, we detach the term $q$. In fact, this omission does not affect convergence. We prove this with the following four steps:

% 1. \Cref{appendix:quantile-regression}  been proved in Appendix I.1 that
% 2. \ref{eqs:valina-quantile-regression} we analyze the original \eq{I.8} 
% 3. \ref{eqs:valina-quantile-regression} the following form is used, which is equivalent to \eq{I.8}
\noindent\textbf{Step 1: Approximation of Quantile Regression Loss.} Since it has been proved in \Cref{appendix:quantile-regression}  that \eq{\ref{eqs:debias-quantile-loss}} is unbiased, we analyze the original \eq{\ref{eqs:valina-quantile-regression}} for simplicity. For theoretical rigor, the following form is used, which is equivalent to \eq{\ref{eqs:valina-quantile-regression}}:
\begin{equation}\label{eq:surropgate-quantile-regression-2}
    \loss_{\text{QR-2}}(\beta;\mathbf{s}_u) := \frac{K+\epsilon}{\vert \itemSet \vert}\beta + \frac{1}{\vert \itemSet \vert}\sum\limits_{i\in\itemSet}(\sui-\beta)_+
\end{equation}
where $\epsilon$  is a sufficiently small constant to ensure the uniqueness of the solution. To bound $q$, we need derive the term $\frac{\partial \beta_u^K}{\partial \mathbf{s}_u}$. However, the non-smooth term $(\cdot)_+$ in \eq{\ref{eq:surropgate-quantile-regression-2}} hinders the derivation. Since $(\cdot)_+$ can be approximated by the softplus function $\phi(x) = \tauQR \ln(1 + \exp(\cdot /\tauQR))$ as $\tauQR \to 0$, we approximate \eq{\ref{eq:surropgate-quantile-regression-2}} with:
\begin{equation}\label{eqs:mu-strongly-convex-approximation-quantile}
\loss_{\text{QR-S}}(\beta;\mathbf{s}_u) \!\!:= \!\!\frac{K+\epsilon}{\vert \itemSet \vert}\beta \!+\! \frac{\mu}{2}\beta^2 \!+ \!\frac{1}{\vert \itemSet \vert}\sum\limits_{i\in\itemSet}\tauQR \ln\left(1+\exp((\sui-\beta)/\tauQR) \right)
\end{equation}
where $\frac{\mu}{2}\beta^2$ ensures its $\mu$-strongly convex. We prove that the difference between $\beta_{u\text{-S}}^K$ (optimal solution of \eq{\ref{eqs:mu-strongly-convex-approximation-quantile}}) and $\margin$ (optimal solution of \eq{\ref{eq:surropgate-quantile-regression-2}}) is negligible, which justifies the use of $\lossB$.

\emph{Proof.} Since $x _+ \le \phi(x) \le x _+ + \tauQR \ln 2$, we have:
\begin{equation}
    \lossA(\beta;\mathbf{s}_u) \le \lossB(\beta;\mathbf{s}_u) \le \lossA(\beta;\mathbf{s}_u) + \frac{\mu}{2} \beta^2 + \epsilon^2 \ln 2 
\end{equation}
Let $A = \frac{\mu}{2}(\margin)^2 + \epsilon^2\ln2$, we have:
\begin{equation}
         \!\!\!-\lossA(\margin;\mathbf{s}_u) \!\le\! -\lossB(\margin;\mathbf{s}_u) + A \le -\lossB(\beta_{u\text{-S}}^K;\mathbf{s}_u) + A
\end{equation}
Thus, we have:
\begin{equation}
\lossA(\beta_{u\text{-S}}^K;\mathbf{s}_u) - \lossA(\margin;\mathbf{s}_u) \!\le\!\lossA(\beta_{u\text{-S}}^K;\mathbf{s}_u) - \lossB(\beta_{u\text{-S}}^K;\mathbf{s}_u) + A
\end{equation}
Since $\lossA(\beta_{u\text{-S}}^K;\mathbf{s}_u) \!- \!\lossB(\beta_{u\text{-S}}^K;\mathbf{s}_u) \!\le\! 0$ and $\margin \!\in\! [-1,1]$, we have:
\begin{equation}
\!\!\!\!\!\lossA(\beta_{u\text{-S}}^K;\mathbf{s}_u) - \lossA(\margin;\mathbf{s}_u) \!\le\! A\!\le\!\frac{\epsilon^2}{2} + \epsilon^2 \ln 2 = \mathcal{O}(\epsilon^2) 
\end{equation}
Considering the sub-gradient of $\mathcal{L} _{\text{QR}-2}$ \wrt $\margin$, i.e., $\partial\lossA(\margin;\mathbf{s}_u) = \Big[\frac{\epsilon - 1}{|\itemSet|}, \frac{\epsilon}{\vert \itemSet \vert}\Big]$, we can analyze the two cases:
\begin{itemize}[leftmargin=*]
    \item Condition: $\beta_{u\text{-S}}^K \ge \margin$, take the sub-gradient $g_1 = \epsilon/\vert\itemSet\vert \ge 0$:
    \begin{equation}
         |\beta_{u\text{-S}}^K - \margin| \le \frac{1}{\vert g _1 \vert}\Big(\lossA(\beta_{u\text{-S}}^K;\mathbf{s}_u) - \lossB(\margin;\mathbf{s}_u) \Big) = \mathcal{O}(\epsilon) 
    \end{equation}
    \item $\beta_{u\text{-S}}^K < \margin$, take the sub-gradient $g_1 = (\epsilon - 1)/\vert\itemSet\vert \ge 0$:
    \begin{equation}
     |\beta_{u\text{-S}}^K - \margin| \le \frac{1}{\vert g _2\vert}\Big(\lossA(\beta_{u\text{-S}}^K;\mathbf{s}_u) - \lossB(\margin;\mathbf{s}_u) \Big) = \mathcal{O}(\epsilon^2) \le \mathcal{O}(\epsilon) 
    \end{equation}
\end{itemize}
In all the cases, we have $|\beta_{u\text{-S}}^K - \margin| \le \mathcal{O}(\epsilon)$, indicating that the surrogate $\lossB$ does not affect the quantile estimation. 

% Since $\kappa \to 0$ in the smooth approximation limit, $\|q\|\to 0$ holds.

\noindent
\textbf{Step 2: The Gradient $q$ bound.} Suppose $\vert \sui - \margin \vert \ge \assCons$ for all $i$, we now bound the gradient norm $\Vert q \Vert$ as:
\[
 \|q\| \;\le\; \frac{2 \vert \itemSet \vert^{1/2}}{\tau}   \frac{e^{-\assCons/\kappa}}{\kappa^2} \quad \text{where} \quad q = \raisebox{-0.25ex}{\colorbox{violet!10}{$\displaystyle \frac{\partial \mathcal{L}_{\text{Talos}}}{\partial \beta_u^K}$}}
\raisebox{-0.6ex}{\colorbox{orange!10}{$\displaystyle \frac{\partial \beta_u^K}{\partial \mathbf{s}_u}$}}
\]
\noindent\textbf{(a) \colorbox{violet!10}{First term.}} The $\frac{\partial \loss_{\text{\lossName}}}{\partial \margin}$ term is bounded, as:
\begin{equation}
    \tau\frac{\partial \loss_{\text{\lossName}}}{\partial \margin} = \sum _{i \in \mathcal{P} _u} \frac{\sigma(\margin - s _{ui})}{|\mathcal{P} _u|} - \sum _{j \in \mathcal{N} _u} \frac{\sigma _\tau(s _{uj} - \margin)}{\sum _{v \in \mathcal{N} _u} \sigma _\tau(s _{uv} - \margin)} \sigma(\margin - s _{uj}) 
\end{equation}
Since sigmoid function satisfies $0 \leq \sigma(x) \leq 1$,  $\Vert\! \frac{\partial \loss_{\text{\lossName}}}{\partial \margin}\!\Vert \le 2/\tau$.

\noindent\textbf{(b) \colorbox{orange!10}{Second term.}} Given by the first-order optimality condition, we have $\frac{\partial \lossB}{\partial \margin} = 0$. Take the gradient \wrt $\mathbf{s}_u$, we have:
\begin{equation}
0 \!\!= \!\!\frac{\partial^2 \lossB}{\partial (\margin)^2} \frac{\partial \margin}{\partial \mathbf{s}_u} + \frac{\partial^2 \lossB}{\partial \margin \partial \mathbf{s}_u}  \to \nabla _{\mathbf{s}_u} \margin \!=\! -\Big[\nabla^2 _{\margin} \lossB\Big]^{-1} \nabla^2 _{\margin,\mathbf{s}_u}\lossB 
\end{equation}
Given $\sigma(x)\sigma(-x) \ge 0$, and set $\mu$ as $\tauQR$, we have:
\begin{equation}
    \nabla _{\margin}^2 \lossB = \mu + \frac{1}{\tauQR |\itemSet|} \sum _{i \in \itemSet} \sigma \Big((\sui-\margin)/\tauQR\Big) \sigma \Big((\margin - \sui)/\tauQR\Big) \ge \tauQR
\end{equation}
We derive second-order gradient as:
\begin{equation}
\nabla _{\margin,\mathbf{s}_u}^2 \lossB = \frac{1}{\tauQR |\itemSet|}\Bigg[\sigma\Big((\sui-\margin)/\tauQR\Big)\sigma\Big((\margin-\sui)/\tauQR\Big)\Bigg] _{i=1}^{|\itemSet|} 
\end{equation}
$\sigma(x)\sigma(-x)$ is an even function that satisfies
$\sigma(x)\sigma(-x) \!=\! \sigma(\vert x \vert)\sigma(\vert -x \vert) \le \exp(\vert x \vert)$. Given $|\sui-\margin|\ge \assCons$, we have $\nabla _{\margin,\mathbf{s}_u}^2 \lossB \le e^{-\assCons /\tauQR}/\tauQR \cdot \mathbf{1}$, where $\mathbf{1}$ is a vector of ones with length $|\itemSet|$. Therefore, we have $ \Vert \nabla _{\mathbf{s}_u} \margin \Vert \le \vert \itemSet \vert^{1/2}\cdot e^{-\assCons/\tauQR}/\tauQR^2$.

\noindent Given the \colorbox{violet!10}{First term} and \colorbox{orange!10}{Second term} bound, we conclude:
\begin{equation}
     \Vert q\Vert \le \frac{2 \vert \itemSet \vert^{1/2}}{\tau}   \frac{e^{-\assCons/\kappa}}{\kappa^2} \implies \Vert q \Vert \to 0 \ \text{as} \ \kappa \to 0
\end{equation}

\begin{table*}[t]
    \centering
    \begin{minipage}[t]{0.49\textwidth}
        \centering
        \captionof{table}{Temporal shift exploration.}
        \label{table: the temporal shift scenarios with SLatK}
        \scriptsize
        \begin{tabular}{l|cc|cc}
            \Xhline{1.3pt}
            \multirow{2}{*}{\textbf{Loss}} & \multicolumn{2}{c|}{\textbf{Gowalla}} & \multicolumn{2}{c}{\textbf{Games}}\bigstrut\\
            \cline{2-5}
            & \textbf{Precision@20} & \textbf{Recall@20} & \textbf{Precision@20} & \textbf{Recall@20}\bigstrut[t]\\
            \Xhline{1.0pt}
            SL@$K$        & 0.0567 & 0.1559 & 0.0091 & 0.0708  \bigstrut\\
            \lossName  & \textbf{0.0574} & \textbf{0.1577} & \textbf{0.0094} & \textbf{0.0738}  \bigstrut[b]\\
            \hline
           \textcolor{red}{\textbf{Imp.\%}} & \textcolor{red}{\textbf{+1.12\%}} & \textcolor{red}{\textbf{+1.12\%}} & \textcolor{red}{\textbf{+2.95\%}} & \textcolor{red}{\textbf{+4.15\%}} \bigstrut\\
            \Xhline{1.3pt}
        \end{tabular}
    \end{minipage}
    \hfill
   \begin{minipage}[t]{0.49\textwidth}
        \centering
        \captionof{table}{Number of negative samples exploration.}
        \label{table: the varying negative numbers with SLatK}
        \scriptsize
        \begin{tabular}{l|ccc|ccc}
            \Xhline{1.3pt}
            \multirow{2}{*}{\textbf{Loss}} & \multicolumn{3}{c|}{\textbf{Beauty}} & \multicolumn{3}{c}{\textbf{MovieLens}}\bigstrut\\
            \cline{2-7}
            & \textbf{$\vert\sampledNegative\vert=8$} & \textbf{$\vert\sampledNegative\vert=16$} & \textbf{$\vert\sampledNegative\vert=32$} & \textbf{$\vert\sampledNegative\vert=8$} & \textbf{$\vert\sampledNegative\vert=16$} & \textbf{$\vert\sampledNegative\vert=32$}\bigstrut\\
            \Xhline{1.0pt}
            SL@$K$     & 0.0167 & 0.0159  & 0.0150  & 0.2242 & 0.2265 & 0.2215  \bigstrut[t]\\
            \lossName  & \textbf{0.0173} & \textbf{0.0162} & \textbf{0.0156} & \textbf{0.2310} & \textbf{0.2311} & \textbf{0.2266}  \bigstrut[b]\\
            \hline
           \textcolor{red}{\textbf{Imp.\%}} & \textcolor{red}{\textbf{+3.67\%}} & \textcolor{red}{\textbf{+1.96\%}} & \textcolor{red}{\textbf{+3.50\%}} & \textcolor{red}{\textbf{+3.01\%}} & \textcolor{red}{\textbf{+2.02\%}} & \textcolor{red}{\textbf{+2.28\%}} \bigstrut\\
            \Xhline{1.3pt}
        \end{tabular}
    \end{minipage}
\end{table*}

\begin{table*}[t]
    \centering
    \begin{minipage}[t]{0.48\textwidth}
        \centering
        \captionof{table}{Performance comparison with varying $K$ on MovieLens. P@$K$ denotes Precision@$K$.}
        \label{table: the varying K with SLatK}
        \scriptsize
        \begin{tabular}{l|cccccc}
            \Xhline{1.3pt}
\multicolumn{1}{c|}{\textbf{Loss}} & \multicolumn{1}{c}{\textbf{P@20}} & \multicolumn{1}{c}{\textbf{P@50}} & \multicolumn{1}{c}{\textbf{P@80}} & \multicolumn{1}{c}{\textbf{P@100}} & \multicolumn{1}{c}{\textbf{P@200}}& \multicolumn{1}{c}{\textbf{P@400}}\bigstrut\\
            \cline{2-7}
            \Xhline{1.0pt}
            SL@$K$     & 0.2287 &0.1557   & 0.1224   & 0.1067  & 0.0685  & 0.0400   \bigstrut[t]\\
            \lossName  & \textbf{0.2349 } & \textbf{0.1600 } & \textbf{0.1257 } & \textbf{0.1101 } & \textbf{0.0700 } & \textbf{0.0410 }  \bigstrut[b]\\
            \hline
           \textcolor{red}{\textbf{Imp.\%}} & \textcolor{red}{\textbf{+2.70\%}} & \textcolor{red}{\textbf{+2.76\%}} & \textcolor{red}{\textbf{+2.65\%}} & \textcolor{red}{\textbf{+3.21\%}} & \textcolor{red}{\textbf{+2.16\%}} & \textcolor{red}{\textbf{+2.26\%}} \bigstrut\\
            \Xhline{1.3pt}
        \end{tabular}
    \end{minipage}
    \hfill
    \begin{minipage}[t]{0.48\textwidth}
    \centering
\caption{The error between the estimated \topk quantile $\hat{\beta}_u^{20}$ and the ideal \topk quantile $\beta_u^{20}$.}
    \label{tab:error-of-quantile-estimation}
\begin{tabular}{l|cccc}
\Xhline{1.3pt}
\multicolumn{1}{c|}{\textbf{Loss}} & \multicolumn{1}{c}{\textbf{Gowalla}} & \multicolumn{1}{c}{\textbf{Beauty}} & \multicolumn{1}{c}{\textbf{Games}} & \multicolumn{1}{c}{\textbf{MovieLens}} \bigstrut\\
\Xhline{1.0pt}
 SL@$K$   &  0.1943
       &       0.1948       &       0.2028       &  0.1389     
                  \bigstrut[t]\\
 \lossName\    &  0.0131       &       0.0123
       &       0.0100
       &       0.0059                             \bigstrut[b]\\
\Xhline{1.3pt}
\end{tabular}
    \end{minipage}
\end{table*}

        % \centering
        % \captionof{table}{Inconsistency between NDCG@$K$ and \topk accuracy. P@20 (R@20) denotes Precision@20 (Recall@20).}
        % \label{table: the inconsistency between SLatK and topk}
        % \scriptsize
        % \begin{tabular}{l|cc|cc|cc|cc}
        %     \Xhline{1.3pt}
        %     \multirow{2}{*}{\textbf{Metric}} & \multicolumn{2}{c|}{\textbf{Gowalla}} & \multicolumn{2}{c|}{\textbf{Games}} & \multicolumn{2}{c|}{\textbf{Beauty}} & \multicolumn{2}{c}{\textbf{MovieLens}}\\
        %     \cline{2-9}
        %     & \textbf{P@20} & \textbf{R@20} & \textbf{P@20} & \textbf{R@20}  & \textbf{P@20} & \textbf{R@20}  & \textbf{P@20} & \textbf{R@20}  \bigstrut\\
        %     \Xhline{1.0pt}
        %     NDCG@$20$     & 19.01\% & 18.24\%   & 21.14\% & 22.21\%   & 22.02\% & 22.55\%    & 17.65\% & 17.51\%    \bigstrut\\
        %     \hline
        %     \Xhline{1.3pt}
        % \end{tabular}

% \begin{table}[t]

% \end{table}

% \begin{table}[t]
% \caption{Quantile estimation error}
% \label{tab:error-of-quantile-estimation}
% \centering

% \begin{tabular}{l|cccc}
% \Xhline{1.3pt}
% \multicolumn{1}{c|}{\textbf{Loss}} & \multicolumn{1}{c}{\textbf{Gowalla}} & \multicolumn{1}{c}{\textbf{Beauty}} & \multicolumn{1}{c}{\textbf{Games}} & \multicolumn{1}{c}{\textbf{MovieLens}} \bigstrut\\
% \Xhline{1.0pt}
%  SL@$K$   &  0.1943
%        &       0.1948       &       0.2028       &  0.1389     
%                   \bigstrut[t]\\
%  \lossName\    &  0.0131       &       0.0123
%        &       0.0100
%        &       0.0059                             \bigstrut[b]\\
% \Xhline{1.3pt}
% \end{tabular}
% \end{table}

\noindent
\textbf{Step 3: Convergence Guarantee.} Let $K$ be the Lipschitz constant of $\loss_{\lossName}$, and $\theta_1^{(t)}$ denotes parameters that are updated at iteration $t$ with $q$ detached. For convenience, we simplify notation $g(\theta_1^{(t)})$ as $g^{(t)}$, $q(\theta_1^{(t)})$ as $q^{(t)}$, and $\loss_{\text{\lossName}}(\theta_1^{(t)})$ as $\loss_{\text{\lossName}}^{(t)}$. We can derive:
\begin{equation}
\begin{aligned}
        \loss_{\text{\lossName}}^{(t+1)} \le& \loss_{\text{\lossName}}^{(t)} - \Big\langle g^{(t)} + q^{(t)}, \alpha g^{(t)}\Big\rangle + \frac{K}{2} \alpha^2\Vert g^{(t)} \Vert^2 \\
        =&\loss_{\text{\lossName}}^{(t)} -\alpha \Big\langle g^{(t)}, q^{(t)} \Big\rangle - (\alpha - \frac{K}{2}\alpha^2) \Vert g^{(t)}\Vert^2 \\
        \le& \loss_{\text{\lossName}}^{(t)} + \alpha \Vert g^{(t)} \Vert \Vert q^{(t)} \Vert - (\alpha - \frac{K}{2}\alpha^2) \Vert g^{(t)}\Vert^2
\end{aligned}
\end{equation} 
Let $c = \alpha - \frac{K}{2}\alpha^2$, take $0 < \alpha < \frac{2}{K}$, we have $c \sum_{t=0}^{T}\Vert g^{(t)} \Vert^2 \le \loss_{\text{\lossName}}^{(0)} + \alpha \sum_{t=0}^{T}\Vert q^{(t)} \Vert \Vert g^{(t)} \Vert$.
With Young inequality, we have: $\alpha \Vert q^{(t)} \Vert \Vert g^{(t)} \Vert$ $\le \frac{c}{2}\Vert g^{(t)} \Vert^2 + \frac{\alpha^2}{2 c}\Vert q^{(t)} \Vert^2$. Therefore, we obtain:
\begin{equation}
    \frac{c}{2} \Vert g^{(t)} \Vert^2 \le \loss_{\text{\lossName}}^{(0)}  + \frac{\alpha^2}{2c} \sum\limits_{t=0}^{T}\Vert q^{(t)} \Vert^2
\end{equation}
We have proved in step 2 that $\Vert q \Vert\to 0$ as $\tauQR \to 0$, thus we conclude:
\begin{equation}
\frac{(2\alpha - K\alpha^2)}{4}\mathbb{E}_t\left[\Vert g^{(t)} \Vert^2  \right]\le   \frac{1}{T}\loss_{\lossName}^{(0)} < \infty
   \implies 
\Vert g^{(T)}\Vert^2 \to 0
\end{equation}

\noindent
\textbf{Step 4: Convergence Equivalence.} We additionally demonstrate that omitting $q$ does not affect convergence to the same solutions: difference between $\theta_1^{(T)}$ (updated with $q$ detach) and $\theta_2^{(T)}$ (updated by $g\! +\! q$) is negligible. Given two update schemes, we can derive:
\[
    \theta_{1}^{(t+1)} = \theta_{1}^{(t)} - \alpha g(\theta_1^{(t)}) \ \  \text{and} \ \ \theta_{2}^{(t+1)} = \theta_{2}^{(t)} - \alpha \Big(g(\theta_2^{(t)})  + q(\theta_2^{(t)}) \Big)
\]
For convenience, we simplify $\Vert \theta_{2}^{(t)} - \theta_1^{(t)} \Vert$ as $B_t$, and $ q(\theta_2^{(t)})$ as $q^{(t)}$. Let $L_g$ denotes the Lipschitz constant of $\nabla \loss_{\text{\lossName}}$, we have:
\begin{equation}
    \begin{aligned}
        B_{t+1} \le& B_{t} + \alpha \Vert g(\theta_2^{(t)}) - g(\theta_1^{(t)}) \Vert + \alpha \Vert q^{(t)} \Vert \\
        \le& B_t + \alpha L_g B_t + \alpha \Vert q^{(t)}\Vert = (1+\alpha L_g) B_t + \alpha \Vert g^{(t)}\Vert
    \end{aligned}
\end{equation}
Note that $B_0 = 0$. Given our conclusion in step 2, we have:
\[
B_{t} \le \alpha \frac{2 \vert \itemSet \vert}{\tau} \frac{e^{-\assCons/\kappa}}{\kappa^2} \sum\limits_{k=0}^{t-1} (1+\alpha L_g)^k = \frac{2 \vert \itemSet \vert}{\tau} \frac{e^{-\assCons/\kappa}}{\kappa^2} \frac{(1+\alpha L_g)^t - 1}{L_g}
\]
Let $T$ denotes the maximum training step. $\forall \epsilon >0$, take a small $\tauQR \ge \sqrt{  2\vert \itemSet \vert\Big((1+\alpha L_g)^T - 1\Big) / \tau\epsilon L_g  }$, we have $\Vert \theta_1^{(T)} - \theta_2^{(T)} \Vert \le \epsilon$, indicating that two update schemes converge to the same optimal solutions.

% \begin{figure}[h]
%     \centering
%     \includegraphics[width=0.8\linewidth]{SLatKPrecisionOnMovieLens/PrecisionSLatK.pdf}
%     \caption{Performance comparisons with varying $K$ on on MF backbone in MovieLens dataset.}
%     \label{fig:SLatK-Varying-K-On-MovieLens}
% \end{figure}

\section{SL@$K$ Comparison}\label[appendix]{appendix-Comparison-with-SL@K}
% \subsection{Comparison between SL@$K$ and TL@$K$}\label{appendix-Comparison-with-SL@K}
% thereby motivating the development of TL@$K$

% We compare SL@$K$~\citep{yang2025breaking} and \lossName\ across multiple evaluation scenarios. We also refer to \citet{yang2025breaking}, and conduct additional comparisons on MovieLens dataset to extend the evaluation scope.

 We present a comparative study between SL@$K$~\citep{yang2025breaking} and \lossName\ across multiple evaluation scenarios. We also refer to \citet{yang2025breaking}, and conduct additional comparisons on MovieLens dataset to extend the evaluation scope.

We first argue that NDCG@$K$ also exhibits significant difference from Top-$K$ accuracy. As shown in \Cref{fig:incosistency-between-topk-metrics}, optimizing NDCG@$K$ may not always yield better \topk accuracy. This observation indicates that SL@$K$, while tailored to NDCG@$K$, may fail to deliver proportional improvements in \topk accuracy. \Cref{table: the varying K with SLatK} also demonstrates this: With varying $K$, \lossName\ consistently outperforms SL@$K$ $+2.62\%$ on average in terms of \topk accuracy metric, underscoring the importance of directly optimizing \topk accuracy in practical recommendation scenarios.

\Cref{table: the varying negative numbers with SLatK} shows that \lossName\ achieves an average improvement of $+2.83\%$, whereas SL@$K$ shows reduced robustness under limited negative numbers. In temporal shift scenario, \lossName\ 
achieves an average improvement of $+2.84\%$ over SL@$K$. 
This superiority reflects the inherent connection between \lossName\ and DRO (\cf \Cref{theorem:DRO}). While prior work~\cite{wu2023understanding} demonstrates the connection between SL and \textit{Distributionally Robustness Optimization}, we attribute the performance drop to the additional weight term introduced in SL@$K$. In addition, \lossName\ incorporates a more accurate threshold estimation strategy, resulting in substantially lower estimation error than SL@$K$ (\cf \Cref{tab:error-of-quantile-estimation}).

\vspace{-5pt}
\section{Detailed Related Work}\label[appendix]{appendix-detailed-related-work}

% \noindent\textbf{Recommendation Models.} In the realm of RS, recommendation models play a vital role in anticipating the user preference. Among various architectures, collaborative filtering (CF)~\cite{su2009survey,zhu2019improving} is widely adopted to instruct the model design. The primary task of CF-models is to predict interactions by assessing the similarity between user and item embeddings. Early research focused on Matrix Factorization (MF) \cite{koren2009matrix}, which decomposes the user-item interaction matrix into latent user and item embedding vectors. This approach act as foundational models like MF \citep{koren2009matrix}, SVD \citep{ bell2007modeling}, and NCF \citep{he2017ncf}, \etc. The more advanced methods, inspired by the efficacy of Graph Neural Networks (GNNs,~\cite{wu2022graph,wang2019neural}), including LightGCN~\cite{he2020lightgcn}, NGCF~\cite{wang2019neural}, LCF~\cite{yu2020graph}, APDA~\cite{zhou2023adaptive}, have emerged and achieve great success to address this issue. Recently, some works attempt to introduce contrastive learning paradigm into LightGCN to augment graph data such as XSimGCL~\cite{yu2023xsimgcl}, $\etc$, achieving SOTA performance. 

In this section, we presnet the detailed related work on \textbf{recommendation models} and \textbf{recommendation loss consistency}.

\noindent\textbf{Recommendation Models.} In the realm of RS, recommendation models play a vital role in anticipating the user preference. Among various architectures, collaborative filtering (CF)~\cite{su2009survey,zhu2019improving,cui2025hatllm} is widely adopted to instruct the model design. The primary task of CF-models is to predict interactions by assessing the similarity between user and item embeddings. Early research focused on Matrix Factorization (MF)~\cite{koren2009matrix}, which decomposes the user-item interaction matrix into latent user and item embedding vectors. This approach act as foundational models like MF~\citep{koren2009matrix}, SVD~\citep{ bell2007modeling} and  NCF~\citep{he2017ncf}. The more advanced methods, inspired by the efficacy of Graph Neural Networks (GNNs~\cite{wu2022graph,wang2019neural,chen2024sigformer,chen2025rankformer}), including LightGCN~\cite{he2020lightgcn}, NGCF~\cite{wang2019neural}, LCF~\cite{yu2020graph}, and APDA~\cite{zhou2023adaptive}, have emerged and achieve great success to address this issue. Recently, some works attempt to introduce contrastive learning paradigm into LightGCN to augment graph data such as XSimGCL~\cite{yu2023xsimgcl}, $\etc$, achieving the state-of-the-art performance. In addition, several recent works utilized large language models~\cite{wang2025msl,wang2025llm4dsr,cui2024distillation} to enhance RS performance.

\noindent\textbf{Recommendation Loss Consistency.} Recent studies~\cite{wydmuch2018no,long2013consistency,pu2025understanding} have theoretically demonstrated the loss consistency in terms of recommendation metrics. In particular, \citet{long2013consistency,wydmuch2018no} have demonstrated H-consistency and Bayes-consistency for SL \wrt Precision@$K$, respectively. \citet{pu2025understanding} further demonstrate the consistency of SL in two-tower recommendation model settings. However, practical recommendation tasks are inherently complex: distribution shifts, sparse interactions, and limited model capacity mean that the Bayes and H-consistency optimal situation is rarely attainable in practice. Consequently, substantial performance disparities among SL-based methods were observed in empirical experiments (\cf \Cref{table: overall performance,table: OOD-performance-comparison}). This necessitates examining their consistency in practical RS scenarios.

\section{Experimental Details}\label[appendix]{appendix:experimental-detials}

\subsection{Inconsistency Simulation Details}\label[appendix]{appendix-simulation-detials}
To quantify the inconsistency between LLPAUC/NDCG and \topk metrics (\ie Recall@$K$ and NDCG@$K$), we simulate pair-wise comparisons of ranking lists as follows: We randomly generate two ranking lists, in which the elements represent the ranking of positive items ($\pi_{ui} \le 200$ for simulating the real-world recommender systems). Afterwards, we compute NDCG, LLPAUC, and \topk metrics (\ie Precision@20, NDCG@20) for both lists. An inconsistency case occurs when one list achieves higher NDCG/LLPAUC but lower Top-K accuracy compared to the other, \ie optimizing LLPAUC/NDCG does not bring benefits for improving \topk metrics. The ratio of such cases is collected over 10,000 independent trials (per dataset) to ensure statistical stability.

\subsection{Recommendation Backbones}\label[appendix]{appendix-recommendation-backbone}
In our experiments, we implement three popular recommendation backbones:
\begin{itemize}[topsep=0pt,leftmargin=10pt]
    \setlength{\itemsep}{0pt}
    \item \textbf{MF} \citep{koren2009matrix}: MF is the most foundational yet effective model that factorizes user-item interactions into learnable embeddings (user embedding and item embedding). All the embedding-based recommendation models use MF as the first layer. Following \citet{wang2019neural}, we implement MF with embedding dimension  $d = 64$ for all settings.
    \item \textbf{LightGCN} \citep{he2020lightgcn}: LightGCN is a effective GNN-based recommendation model that aggregates high-order user-item interactions via non-parameterized graph convolution.
    By eliminating nonlinear activations and feature transformations in NGCF \citep{wang2019neural}, LightGCN achieves computational efficiency and enhanced performance. In our experiments, we adopt 2 graph convolution layers, which aligns with the original setting in \citet{he2020lightgcn}.
    \item \textbf{XSimGCL} \citep{yu2023xsimgcl}: XSimGCL is a novel contrastive learning enhanced~\citep{jaiswal2020survey, liu2021self} variant of 3-layer LightGCN that injects random noise into intermediate embeddings, and optimizes an auxiliary InfoNCE loss \citep{oord2018representation} between the final layer and a selected intermediate layer ($l^*$).
    Following the original \citet{yu2023xsimgcl}'s setting, the modulus of random noise between each layer is set as 0.1, the contrastive layer $l^*$ is set as 1 (where the embedding layer is 0-th layer), the temperature of InfoNCE is set as 0.1, and the weight of the auxiliary InfoNCE loss is searching from $\{0.01, 0.05, 0.1, 0.2\}$.
\end{itemize}

\begin{table}[t]
\caption{Dataset statistics.}
\label{tab:dataset statistics}
\centering

\begin{tabular}{l|rrrr}
\Xhline{1.3pt}
\multicolumn{1}{c|}{\textbf{Dataset}} & \multicolumn{1}{c}{\textbf{\#Users}} & \multicolumn{1}{c}{\textbf{\#Items}} & \multicolumn{1}{c}{\textbf{\#Interactions}} & \multicolumn{1}{c}{\textbf{Density}} \bigstrut\\
\Xhline{1.0pt}
 Electronics     & 150,523   & 52,024   & 1,312,545        & 0.0007                   \bigstrut[t]\\
 Gowalla        & 29,858   & 40,988   & 1,027,464        & 0.0007                   \\
 Games         & 18,813   & 8,691    & 177,572       & 0.0009                      \\
 Beauty         & 15,603   & 8,693   & 139,554        & 0.0009                      \bigstrut[b]\\

\Xhline{1.3pt}
\end{tabular}

\end{table}

\subsection{Datasets}\label[appendix]{appendix-dataset}
The four benchmark datasets used in our experiments are summarized in \Cref{tab:dataset statistics}. Following~\citep{he2016vbpr,wang2019neural}, we use 10-core setting (or 5-core setting) for Gowalla (or Amazon) dataset. Following~\citet{yang2024psl}, we further clean the data by excluding interactions with ratings below 3 (if available). The dataset is randomly split into training set, validation set, and test set in a ratio of 7:1:2. The details of datasets are as follows:
\begin{itemize}[topsep=0pt,leftmargin=10pt]
    \item \textbf{Gowalla}~\citep{he2020lightgcn}: A check-in dataset from the location-based social network Gowalla\footnote{\textcolor{magenta}{\url{https://en.wikipedia.org/wiki/Gowalla}}}, which consists of 1M users, 1M locations, and 6M check-ins. 

    \item \textbf{Movielens} \citep{harper2015movielens}: The Movielens dataset is a movie rating dataset collected from Movielens\footnote{\url{https://movielens.org/}}. We use the Movielens-100K version, which contains 100,000 ratings from 1000 users on 1700 movies.

    \item \textbf{Amazon:} Subsets of the 2014 Amazon product review 
    corpus\footnote{\textcolor{magenta}{\url{https://cseweb.ucsd.edu/~jmcauley/datasets/amazon/links.html}}}, which contains 142.8 million reviews spanning May 1996 to July 2014. We process three widely-used categories: Beauty, Games and Electronics~\cite{yang2024psl}, with interactions ranging from 40K to 1M.
\end{itemize}

% Specifically, for each dataset, we randomly sample ranking list pairs and calculate the ratio of inconsistency cases where one list outperforms the other in NDCG but underperforms it in \topk accuracy (\eg Precision@20). 

\begin{table*}[t]
\centering
\caption{Overall performance comparison of \lossName\ with other losses. \colorbox{lightblue}{\textcolor{customblue}{blue}} indicates the \lossName\ achieves the SOTA performance, and the runner-up is underlined. \textcolor{customred}{Imp.\%} indicates the relatively improvements of \lossName\ over the best baselines. The mark `*' suggests the improvement is statistically significant with $p<0.05$. }
\label{table: overall performance on MRR and NDCG}
\resizebox{\textwidth}{!}{
\begin{tabular}{c|l|cc|cc|cc|cc} 
\Xhline{1.5pt}
\multirow{2}{*}{\textbf{Model}} & \multicolumn{1}{c|}{\multirow{2}{*}{\textbf{Loss}}} & \multicolumn{2}{c|}{\textbf{Gowalla}}                           & \multicolumn{2}{c|}{\textbf{Beauty}}                     & \multicolumn{2}{c|}{\textbf{Games}}                 & \multicolumn{2}{c}{\textbf{Electronics}}                      \bigstrut[t]\\ 
\cline{3-10}
& \multicolumn{1}{c|}{}  & \multicolumn{1}{c}{\textbf{MRR@20}} & \multicolumn{1}{c|}{\textbf{NDCG@20}}   & \multicolumn{1}{c}{\textbf{MRR@20}} & \multicolumn{1}{c|}{\textbf{NDCG@20}}  & \multicolumn{1}{c}{\textbf{MRR@20}} & \multicolumn{1}{c|}{\textbf{NDCG@20}}   & \multicolumn{1}{c}{\textbf{MRR@20}} & \multicolumn{1}{c}{\textbf{NDCG@20}}  \bigstrut[t]\\ 
\Xhline{1.0pt}
\multirow{14}{*}{MF}      
&BPR&$0.0340$&$0.1167$&$0.0374$&$0.0698$&$0.0339$&$0.0699$&$0.0155$&$0.0303$\bigstrut[t]\\
&AATP&$0.0227$&$0.0795$&$0.0294$&$0.0558$&$0.0318$&$0.0674$&$0.0065$&$0.0135$ \\
&RS@$K$&$0.0301$&$0.0972$&$0.0223$&$0.0374$&$0.0232$&$0.0458$&$0.0050$&$0.0088$ \\
&SmoothI@$K$&$0.0449$&$0.1473$&$0.0443$&$0.0778$&$0.0492$&$0.0973$&$0.0180$&$0.0329$ \\
&SL&$0.0481$&$0.1585$&$0.0470$&$0.0813$&$0.0511$&$0.1027$&$0.0175$&$0.0339$ \\
&BSL&$0.0481$&$0.1585$&\underline{$0.0470$}&$0.0813$&\underline{$0.0514$}&$0.1035$&$0.0177$&$0.0341$ \\
&PSL&\underline{$0.0485$}&\underline{$0.1595$}&$0.0469$&\underline{$0.0820$}&$0.0514$&\underline{$0.1035$}&\underline{$0.0183$}&\underline{$0.0350$} \\
&AdvInfoNCE&$0.0481$&$0.1582$&$0.0468$&$0.0811$&$0.0505$&$0.1027$&$0.0180$&$0.0341$ \\
&LLPAUC&$0.0444$&$0.1459$&$0.0395$&$0.0737$&$0.0450$&$0.0922$&$0.0134$&$0.0285$ \\
&\lossName&\cellcolor{lightblue}\textcolor{customblue}{$\textbf{0.0513}$}&\cellcolor{lightblue}\textcolor{customblue}{$\textbf{0.1667}$}&\cellcolor{lightblue}\textcolor{customblue}{$\textbf{0.0482}$}&\cellcolor{lightblue}\textcolor{customblue}{$\textbf{0.0851}$}&\cellcolor{lightblue}\textcolor{customblue}{$\textbf{0.0532}$}&\cellcolor{lightblue}\textcolor{customblue}{$\textbf{0.1063}$}&\cellcolor{lightblue}\textcolor{customblue}{$\textbf{0.0190}$}&\cellcolor{lightblue}\textcolor{customblue}{$\textbf{0.0361}$}\bigstrut[b]\\
\hhline{~*{9}{!{\arrayrulecolor{black}}-}} % 黑色虚线
&\textcolor{customred}{\textbf{Imp.\%}}&$\cellcolor{pink!20}\textcolor{customred}{\textbf{+5.84\%}^*}$&$\cellcolor{pink!20}\textcolor{customred}{\textbf{+4.49\%}^*}$&$\cellcolor{pink!20}\textcolor{customred}{\textbf{+2.39\%}^*}$&$\cellcolor{pink!20}\textcolor{customred}{\textbf{+3.77\%}^*}$&$\cellcolor{pink!20}\textcolor{customred}{\textbf{+3.35\%}^*}$&$\cellcolor{pink!20}\textcolor{customred}{\textbf{+2.69\%}^*}$&$\cellcolor{pink!20}\textcolor{customred}{\textbf{+3.63\%}^*}$&$\cellcolor{pink!20}\textcolor{customred}{\textbf{+3.15\%}^*}$\bigstrut\\
\Xhline{1.0pt}
\multirow{14}{*}{LGCN}      
&BPR&$0.0425$&$0.1398$&$0.0412$&$0.0762$&$0.0467$&$0.0952$&$0.0128$&$0.0251$\bigstrut[t]\\
&AATP&$0.0153$&$0.0533$&$0.0290$&$0.0560$&$0.0262$&$0.0574$&$0.0069$&$0.0144$ \\
&RS@$K$&$0.0392$&$0.1321$&$0.0410$&$0.0695$&$0.0414$&$0.0816$&$0.0103$&$0.0195$ \\
&SmoothI@$K$&$0.0444$&$0.1472$&$0.0456$&$0.0811$&$0.0496$&$0.0981$& $0.0196$&\underline{$0.0362$} \\
&SL&$0.0480$&$0.1584$&$0.0458$&$0.0810$&\underline{$0.0514$}&\underline{$0.1035$}&$0.0176$&$0.0340$ \\
&BSL&$0.0480$&$0.1584$&$0.0458$&$0.0810$&$0.0511$&$0.1031$&$0.0175$&$0.0338$ \\
&PSL&\underline{$0.0490$}&\underline{$0.1608$}&\underline{$0.0462$}&\underline{$0.0813$}&$0.0514$&$0.1031$&$0.0179$&$0.0343$ \\
&AdvInfoNCE&$0.0481$&$0.1587$&$0.0458$&$0.0811$&$0.0513$&$0.1032$&$0.0175$&$0.0338$ \\
&LLPAUC&$0.0415$&$0.1367$&$0.0447$&$0.0803$&$0.0505$&$0.1012$&$0.0179$&$0.0346$ \\
&\lossName&\cellcolor{lightblue}\textcolor{customblue}{$\textbf{0.0517}$}&\cellcolor{lightblue}\textcolor{customblue}{$\textbf{0.1675}$}&\cellcolor{lightblue}\textcolor{customblue}{$\textbf{0.0478}$}&\cellcolor{lightblue}\textcolor{customblue}{$\textbf{0.0848}$}&\cellcolor{lightblue}\textcolor{customblue}{$\textbf{0.0528}$}&\cellcolor{lightblue}\textcolor{customblue}{$\textbf{0.1056}$}& $0.0190$ &\cellcolor{lightblue}\textcolor{customblue}{$\textbf{0.0363}$}\bigstrut[b]\\
\hhline{~*{9}{!{\arrayrulecolor{black}}-}} % 黑色虚线
&\textcolor{customred}{\textbf{Imp.\%}}&$\cellcolor{pink!20}\textcolor{customred}{\textbf{+5.66\%}^*}$&$\cellcolor{pink!20}\textcolor{customred}{\textbf{+4.17\%}^*}$&$\cellcolor{pink!20}\textcolor{customred}{\textbf{+3.55\%}^*}$&$\cellcolor{pink!20}\textcolor{customred}{\textbf{+4.20\%}^*}$&$\cellcolor{pink!20}\textcolor{customred}{\textbf{+2.71\%}^*}$&$\cellcolor{pink!20}\textcolor{customred}{\textbf{+2.00\%}^*}$&$ -$&$\cellcolor{pink!20}\textcolor{customred}{\textbf{+0.13\%}^*}$\bigstrut\\
\Xhline{1.0pt}
\multirow{14}{*}{XSimGCL}      
&BPR&$0.0469$&$0.1531$&$0.0447$&$0.0812$&$0.0483$&$0.0976$&$0.0177$&$0.0339$\bigstrut[t]\\
&AATP&$0.0311$&$0.1108$&$0.0396$&$0.0707$&$0.0398$&$0.0833$&$0.0134$&$0.0264$ \\
&RS@$K$&$0.0364$&$0.1251$&$0.0401$&$0.0697$&$0.0402$&$0.0824$&$0.0100$&$0.0195$ \\
&SmoothI@$K$&$0.0379$&$0.1299$&$0.0181$&$0.0312$&$0.0330$&$0.0684$&$0.0118$&$0.0224$ \\
&SL&$0.0475$&$0.1568$&$0.0457$&$0.0805$&$0.0503$&$0.1017$&$0.0175$&$0.0338$ \\
&BSL&$0.0475$&$0.1572$&$0.0438$&$0.0788$&$0.0504$&$0.1023$&$0.0168$&$0.0330$ \\
&PSL&$0.0480$&\underline{$0.1583$}&\underline{$0.0463$}&$0.0808$&$0.0505$&$0.1019$&\underline{$0.0181$}&\underline{$0.0348$} \\
&AdvInfoNCE&$0.0475$&$0.1568$&$0.0456$&$0.0803$&$0.0502$&$0.1019$&$0.0175$&$0.0337$ \\
&LLPAUC&\underline{$0.0481$}&$0.1571$&$0.0459$&\underline{$0.0820$}&\underline{$0.0517$}&\underline{$0.1037$}&$0.0181$&$0.0347$ \\
&\lossName&\cellcolor{lightblue}\textcolor{customblue}{$\textbf{0.0506}$}&\cellcolor{lightblue}\textcolor{customblue}{$\textbf{0.1645}$}&\cellcolor{lightblue}\textcolor{customblue}{$\textbf{0.0489}$}&\cellcolor{lightblue}\textcolor{customblue}{$\textbf{0.0858}$}&\cellcolor{lightblue}\textcolor{customblue}{$\textbf{0.0526}$}&\cellcolor{lightblue}\textcolor{customblue}{$\textbf{0.1057}$}&\cellcolor{lightblue}\textcolor{customblue}{$\textbf{0.0189}$}&\cellcolor{lightblue}\textcolor{customblue}{$\textbf{0.0359}$}\bigstrut[b]\\
\hhline{~*{9}{!{\arrayrulecolor{black}}-}} % 黑色虚线
&\textcolor{customred}{\textbf{Imp.\%}}&$\cellcolor{pink!20}\textcolor{customred}{\textbf{+5.18\%}^*}$&$\cellcolor{pink!20}\textcolor{customred}{\textbf{+3.88\%}^*}$&$\cellcolor{pink!20}\textcolor{customred}{\textbf{+5.47\%}^*}$&$\cellcolor{pink!20}\textcolor{customred}{\textbf{+4.60\%}^*}$&$\cellcolor{pink!20}\textcolor{customred}{\textbf{+1.84\%}^*}$&$\cellcolor{pink!20}\textcolor{customred}{\textbf{+1.97\%}^*}$&$\cellcolor{pink!20}\textcolor{customred}{\textbf{+4.15\%}^*}$&$\cellcolor{pink!20}\textcolor{customred}{\textbf{+3.25\%}^*}$\bigstrut\\
\Xhline{1.5pt}
\end{tabular}
}
\end{table*}

\subsection{Recommendation Scenarios}\label[appendix]{appendix-recommendation-scenarios}

The detailed dataset constructions in IID and distributional shift settings are as follows:
\begin{itemize}[topsep=0pt,leftmargin=10pt]
    \item \textbf{IID setting.} Following the standard recommendation setup~\citep{he2020lightgcn}, we \iid split training and test sets from the complete dataset, maintaining identical distributions between both sets. Specifically, the positive items of each user are split into 80\% training and 20\% test sets. Moreover, the training set is further split into 90\% training and 10\% validation sets for hyperparameter tuning.    
    \item \textbf{Distributional shift setting.} We follow~\cite{wang2024distributionally}, introducing temporal bias to construct the test scenario with distribution shifts --- we divide the training and test dataset according to the interaction time. Specifically, interactions with timestamps in the earliest 80\% constitute the training set, while the latest 20\% form the test set. Additionally, we randomly split 10\% of the training set as the validation set.  The temporal shift is very common in real recommendation systems, as user preferences typically evolve over time~\citep{wang2022causal}.
\end{itemize}

\subsection{Hyperparameter Setting}\label[appendix]{appendix-hyperparameter-setting}
Following~\cite{wu2024bsl}, the latent embedding size is set as $64$. For model training, we adopt Adam~\cite{kingma2014adam} optimizer with the learning rate in the range $\{10^{-1}, 10^{-2}, 10^{-3}\}$, except for BPR which uses an extended search $\text{lr} \in \{10^{-1}, 10^{-2}, 10^{-3}, 10^{-4}\}$. The weight decay (\textbf{wd}) is searched in $\{0, 10^{-4}, 10^{-6}, 10^{-8}\}$, except for BPR which uses an extended search $\text{wd} \in \{0, 10^{-3}, 10^{-4}, 10^{-5}, 10^{-6}, 10^{-8}\}$. The learning rate of the quantile estimation in \lossName~is fixed as $10^{-3}$. Compared with SL, \lossName\ preserves simplicity in hyperparameter tuning, requiring only a single temperature parameter $\tau$. On XSimGCL backbone, we follow~\citep{yu2023xsimgcl} tune the weight of the auxiliary InfoNCE (\textbf{wl}) in $\{0.2,0.1,0.05,0.01\}$. Following \citet{he2020lightgcn}, we adopt early stopping strategy, in which the training stops if Precision@20 metric fails to improve for 25 consecutive epochs in validation set. Following \citep{wu2024bsl}, we uniformly sample 1024 negative items for each positive instance in training (BPR only samples one). 
% \textbf{Optimizer.}  We optimize all models using Adam~\citep{kingma2014adam} with a batch size 1024. The learning rate (\textbf{lr}) is searched from $\{10^{-1}, 10^{-2}, 10^{-3}\}$, except for BPR which uses an extended search $\text{lr} \in \{10^{-1}, 10^{-2}, 10^{-3}, 10^{-4}\}$. The weight decay (\textbf{wd}) is searched in $\{0, 10^{-4}, 10^{-6}, 10^{-8}\}$, except for BPR which uses an extended search $\text{wd} \in \{0, 10^{-3}, 10^{-4}, 10^{-5}, 10^{-6}, 10^{-8}\}$. On XSimGCL backbone, we follow~\citep{yu2023xsimgcl} tune the weight of the auxiliary InfoNCE (\textbf{wl}) in $\{0.2,0.1,0.05,0.01\}$. The  Following \citet{he2020lightgcn}, we adopt early stopping strategy, in which the training stops if Precision@20 metric fails to improve for 25 consecutive epochs in validation set. Following the negative sampling strategy in \citet{wu2024bsl}, we uniformly sample 1024 negative items for each positive instance in training.

\noindent\textbf{Hyperparameter Settings.}
To maintain the fairness across all methods, we tune each methods with a very fine granularity to ensure their optimal performance. We reproduced the following losses as baselines in our experiments:
\begin{itemize}[topsep=0pt,leftmargin=10pt]
    \item \textbf{BPR:} No other hyperparameters.

    \item \textbf{AATP:} The quantile regression learning rate:
    
    $\text{lr}_{\text{quantile}} \in \{10^{-1}, 10^{-2}, 10^{-3}, 10^{-4}\}$.

    \item \textbf{RS@$K$:} The temperature $\tau_1$, which approximates postive ranking, is searched in the range $\{0.01, 0.05, 0.10, 0.15, 0.20,0.25,0.30\}$. The temperature $\tau_2$, which approximates the Heaviside function, is searched in $\{1,2,3,4,5\}$.

    \item \textbf{SL:} The temperature $\tau \in \{0.05, 0.10, 0.15, 0.2, 0.25, 0.30\}$.

    \item \textbf{SmoothI@$K$:} The temperature $\tau$ is searched in the same space as SL. Following the original setting~\citet{SmoothIAAAI2021} offset $\delta$ is searched in the rang $\{0.05,0.1,0.15,0.2,0.25,0.30\}$.

    \item \textbf{BSL:} The temperatures $\tau_1, \tau_2$ for positive and negative terms are searched in the same space as SL, respectively.
    
    \item \textbf{PSL:} The temperature $\tau$ is searched in the same space as SL. Following~\citet{yang2024psl}, the activation function is used as $\sigma(\cdot) = \tanh(\cdot / 2)$ for its uniformly SOTA performance.

    \item \textbf{AdvInfoNCE:} The temperature $\tau$ is searched in the same space as SL. Following the original setting~\citet{zhang2024empowering}, the other hyperparameters including: the negative weight is set as $64$, the adversarial learning interval  $T_{\text{adv}}$  is searched in $\{5, 10, 15, 20\}$, the total adversarial training times $E_{\text{adv}}$ is searched in $\{ 5, 10, 15, 20, 25, 30\}$, the adversarial learning rate is searched in $\{10^{-4}, 10^{-5}\}$.

    \item \textbf{LLPAUC:}  hyperparameters $\alpha \in \{0.1, 0.3, 0.5, 0.7, 0.9\}$ and $\beta \in \{0.01, 0.1\}$, which follows \citet{shi2024lower}'s setting.

    \item \textbf{\lossName:} The temperatures $\tau$ are searched in the same space as SL. The learning rate of quantile regression is fixed as $10^{-3}$. No additional hyperparameters.
\end{itemize}
For all compared methods, we closely follow configurations in their respective publications to ensure the optimal performance. All experiments are conducted on one NVIDIA GeForce RTX 4090.

\section{NDCG and MRR performance}\label[appendix]{appendix-on-MRR-and-NDCG}
\noindent\textbf{NDCG and MRR performance in \Cref{table: overall performance on MRR and NDCG}.} Since comparable trends were observed across different backbones, we only report MF results for brevity. While \lossName\ targets optimizing \topk accuracy, it also exhibits superiority on MRR@$K$ and NDCG@$K$. The reason can be attributed to the close relations between these \topk metrics, where NDCG@$K$ and MRR@$K$ is built on Precision@$K$.

\section{Training Efficiency}\label[appendix]{appendix-computational-efficiency}

We evaluate the computational efficiency of \lossName\ on matrix factorization (MF) models and compare it with baseline losses. \Cref{table:computational-efficiency} provides the time complexity and practical computational time of \lossName\ and baseline methods. As shown in \Cref{table:computational-efficiency}, \lossName\ incurs comparable overhead to SL (1.00–1.50$\times$ runtime), aligning with our complexity analysis in \Cref{sec:analyse-on-our-loss}.

\begin{table}[ht]
    \centering
    \scriptsize
    \caption{Computational time (s / epoch) on MF. All methods adopt uniform negative sampling technique, with $\vert\sampledNegative\vert = 1024$.}
    \label{table:computational-efficiency}
    \begin{tabular}{@{}llccccc@{}}
        \Xhline{1.3pt}
        \textbf{Loss} & \textbf{Complexity} & \textbf{Games} & \textbf{Beauty} & \textbf{Gowalla} & \textbf{Electronics} \bigstrut\\ 
        \Xhline{1.0pt}
        BPR         & $\mathcal{O}(\bar{P}\vert\userSet\vert\vert\hat{\sampledNegative}\vert)$ &  1.23       &       0.16       &       0.14       &       2.18            \bigstrut\\
        AATP        & $\mathcal{O}(\bar{P}\vert\userSet\vert\vert\hat{\sampledNegative}\vert)$ &  5.38       &       0.54       &       0.68       &       7.84            \bigstrut\\

        RS@$K$      & $\mathcal{O}(\bar{P}\vert\userSet\vert\vert\hat{\sampledNegative}\vert)$ &  4.58       &       0.43       &       0.37       &       7.30            \bigstrut\\

        SmoothI@$K$     & $\mathcal{O}(\bar{P}\vert\userSet\vert\vert\hat{\sampledNegative}\vert K)$ &  7.13       &       0.93       &       0.63       &       10.54           \bigstrut\\

        SL          & $\mathcal{O}(\bar{P}\vert\userSet\vert\vert\hat{\sampledNegative}\vert)$ &  4.79       &       0.42       &       0.31       &       7.01            \bigstrut\\
        BSL         & $\mathcal{O}(\bar{P}\vert\userSet\vert\vert\hat{\sampledNegative}\vert)$ &  4.56       &       0.43       &       0.34       &       7.17            \bigstrut\\
        PSL         & $\mathcal{O}(\bar{P}\vert\userSet\vert\vert\hat{\sampledNegative}\vert)$ &  4.96       &       0.42       &       0.35       &       7.15            \bigstrut\\

        AdvInfoNCE  & $\mathcal{O}(\bar{P}\vert\userSet\vert\vert\hat{\sampledNegative}\vert)$ &  8.14       &       0.71       &       1.00       &       9.31            \bigstrut\\
        LLPAUC      & $\mathcal{O}(\bar{P}\vert\userSet\vert\vert\hat{\sampledNegative}\vert)$ &  4.82       &       0.52       &       0.37       &       6.63            \bigstrut\\
        \lossName      & $\mathcal{O}(2\bar{P}\vert\userSet\vert\vert\hat{\sampledNegative}\vert)$ &  5.37       &       0.54       &       0.43       &       7.98            \bigstrut\\
        \Xhline{1.3pt}
    \end{tabular}
\end{table}

\section{Notations}
We summarize the notations used in this paper as follows: 
\begin{itemize}[leftmargin=*]
    \item \Cref{tab:notation-for-talos-derivation} provides notations that are used to drive the \lossName\ loss function. 
    \item \Cref{tab:notation-for-talos-dro-property} provides notations that are used to demonstrate the connections between \lossName\ and \textit{Distributional Robustness Optimization}. 
    \item \Cref{tab:notation-for-talos-convergence} provides notations that are used to demonstrate the convergence property of \lossName\ and its corresponding proof.
    \item \Cref{tab:notation-for-quantile-regression} provides notations that are used to demonstrate the vanilla quantile regression and the proof in terms of our unbiased quantile regression loss.
\end{itemize}

\begin{table}[t]
    \centering
        \caption{Notations for deriving \lossName.}
    \begin{tabular}{ll}
\toprule
        \textbf{Notations} & \textbf{Descriptions} \\
        \midrule
         $u$ & a user in the user set $\userSet$\\
         $i$ & an item in the item set $\itemSet$\\
         $\posU$ & the positive item set of user $u$ \\
         $\negU$ & the negative item set of user $u$ \\
         $\sampledNegative$ & the sampled negative item set of user $u$ \\
         $\sui$ & the similarity score between item $i$ and user $u$ \\
         $\pi_{ui}$ & the ranking of item $i$ in ranking list of user $u$ \\
         $\margin$ & the \topk quantile of user $u$ \\
         $\hat{\beta}_u$ & the estimated \topk quantile of user $u$ \\
         $\mathcal{Q}_K(u)$ & the unbiased quantile estimation loss \\
         $\loss_{\text{\lossName}}$ & the \lossName\ loss \\
         $\mathbb{I}(\cdot)$ & the Indicator function \\
         $\delta(\cdot)$ & the Heaviside step function \\
         $\sigma$ & the $\text{sigmoid}(\cdot)$ activation function \\
         $\sigma_\tau$ & the $\text{sigmoid}(\cdot)^{1/\tau}$ activation function \\
    \bottomrule
    \end{tabular}
    \label{tab:notation-for-talos-derivation}
\end{table}

\begin{table}[t]
    \centering
        \caption{Notations for \Cref{theorem:DRO} and its Proof}
    \begin{tabular}{ll}
\toprule
        \textbf{Notations} & \textbf{Descriptions} \\
        \midrule
         $\mathbb{Q}$    & the uncertainty set \\
         $\hat{Q}$       & the uniform distribution of negative item set $\negU$ \\
         $Q$             & the pertubed distribution of $\hat{Q}$ \\
    \bottomrule
    \end{tabular}
    \label{tab:notation-for-talos-dro-property}
\end{table}

\begin{table}[t]
    \centering
        \caption{Notations for \Cref{theorem:convergence} and its Proof}
    \begin{tabular}{ll}
\toprule
        \textbf{Notations} & \textbf{Descriptions} \\
        \midrule
        $\mathbf{s}_u$          & the vector of all predicted scores of user $u$ \\
        $\nabla_{\mathbf{s}_u}$ & the \lossName gradient \wrt $\mathbf{s}_u$     \\
        $\nabla_{\margin}$      & the \lossName gradient \wrt $\margin$          \\
        $g$                     & the abbreviation for $\frac{\partial \mathcal{L}_{\text{Talos}}}{\partial \mathbf{s}_u}$ term\\
        $q$                     & the abbreviation for $\frac{\partial \mathcal{L}_{\text{Talos}}}{\partial \beta_u^K} \frac{\partial \beta_u^K}{\partial \mathbf{s}_u}$ term\\
        $\epsilon$              &  the sufficiently small constant                          \\
        $\kappa$                &  the hyperparameter in softplus function                  \\
        $\loss_{\text{QR-2}}$   & the equivalent form of quantile regression loss           \\
        $\loss_{\text{QR-S}}$   & the convex appximiation form of $\loss_{\text{QR-2}}$     \\
        $\beta_{u\text{-S}}^K$  & the optimal solution of $\loss_{\text{QR-S}}$             \\
        $T$                     & the total optimization step                               \\
        $\theta_1^{T}$          & the model parameter updated with $g$                      \\
        $\theta_2^{T}$          & the model parameter updated with $g+q$                    \\
        $\alpha$                & the fixed step-size in gradient decent                    \\
        $L_g$                   & the Lipschiz constant of $\nabla\loss_{\text{\lossName}}$ \\
    \bottomrule
    \end{tabular}
    \label{tab:notation-for-talos-convergence}
\end{table}

\begin{table}[t]
    \centering
        \caption{Notations for Quantile Regression}
    \begin{tabular}{ll}
\toprule
        \textbf{Notations} & \textbf{Descriptions} \\
        \midrule
        $U(\itemSet)$ & the uniform distribution of item set $\itemSet$    \\
        $S$   & the random variable of the score of items $\sui$ \\
        $F_S$ & the cumulative distribution function \wrt $S$              \\
    \bottomrule
    \end{tabular}
    \label{tab:notation-for-quantile-regression}
\end{table}

\section{Quantile Regression}\label[appendix]{appendix:quantile-regression}

% Quantile estimation~\cite{koenker2005quantile,hao2007quantile,shao2008mathematical} has been well-studied in statistical learning literature. A prominent method is quantile regression~\cite{koenker2005quantile}, which estimates the quantile through the following loss function:
% \begin{equation}
%     \loss_{\mathcal{Q}} = \frac{1}{\vert \itemSet \vert} \sum\limits_{i\in\itemSet} \rho(\sui-\beta,K)
% \end{equation}
% The function $\rho$ is defined as:
% \begin{equation} \label{eqs:quantile-function}
%     \rho(x, K) = -\frac{K}{\vert \itemSet \vert}\cdot(x)_- + \left(1 - \frac{K}{\vert \itemSet\vert}\right)\cdot(x)_+
% \end{equation}
% where $(x)_+ = \max\{0,x\}$, and $(x)_- = \min\{0,x\}$. The convex nature~\cite{koenker2005quantile} of this loss function ensures efficient gradient-based optimization for \topk quantile estimation.

Quantile regression method \citep{koenker2005quantile,hao2007quantile} is uniformly used for sample quantile estimation. The \topk quantile is estimated by the following loss:
\begin{equation}\label{eqs:valina-quantile-regression}
    \loss_{\text{QR}}(u) = \frac{1}{\vert \itemSet \vert} \sum\limits_{i\in\itemSet} \Big((1-\frac{K}{\vert\itemSet\vert})(\sui-\hat\beta_u)_+ + \frac{K}{\vert\itemSet\vert}(\hat\beta_u-\sui)_+\Big)
\end{equation}
or equivalently:
\begin{equation} \label{eq:quantile_regression}
    \mathcal{L}_{\text{QR}}(u) = \mathbb{E}_{i \sim U(\mathcal{I})} \left[ (1 - \frac{K}{\vert\itemSet\vert})(s_{ui} - \hat \beta_u)_+ + \frac{K}{\vert\itemSet\vert}(\hat \beta_u - s_{ui})_+ \right]
\end{equation}
where $(\cdot)_+ = \max(\cdot, 0)$, and $U(\itemSet)$ denotes the uniform distribution on item set $\itemSet$. The \topk quantile is estimated as:
\begin{equation}
 \margin = \mathop{\arg\min}\limits_{\hat{\beta}_u}\loss_{\text{QR}}(u)   
\end{equation}
We show this as follows:
\begin{proof}
    Suppose that $S$ is a random variable representing the score of items $s_{ui}$, and $F_S$ is the \cdf of $S$ on $\mathbb{R}$. Since $i \sim \mathcal{I}$ is uniformly distributed, the quantile regression loss (\ref{eq:quantile_regression}) can be rewritten as the following expectation:
\begin{equation} \label{eq:quantile_regression-3}
    \begin{aligned}
        \mathcal{L}_{\text{QR}}(u) 
        &= \mathbb{E}_{S \sim F_S} \left[ (1 -  \frac{K}{\vert\itemSet\vert})(S - \hat \beta_u)_+ +  \frac{K}{\vert\itemSet\vert}(\hat \beta_u - S)_+ \right]    \\
        &=  \int_{\hat \beta_u}^{\infty} (1 -  \frac{K}{\vert\itemSet\vert})(S - \hat \beta_u) \diff F_S(S) + \int_{-\infty}^{\hat \beta_u}  \frac{K}{\vert\itemSet\vert}(\hat \beta_u - S) \diff F_S(S)
    \end{aligned}
\end{equation}
Compute the  derivative of $\loss_{\text{QR}}(u)$ with respect to $\hat\beta_u$, set it to 0, and let $\beta_u = \mathop{\arg\min}\limits_{\hat\beta_u} \mathcal{L}_{\text{QR}}(u)$, we have:
\begin{equation} \label{eq:quantile_regression-solution}
     \frac{K}{\vert\itemSet\vert} \int_{-\infty}^{\beta_u} \diff F_S(S) = (1 -  \frac{K}{\vert\itemSet\vert}) \int_{\beta_u}^{\infty} \diff F_S(S)
\end{equation}
Given $F_S$ is the \cdf of $S$ on $\mathbb{R}$, we have:
\begin{equation}\label{eq:quantile_regression_cdf}
    F_S(\infty) = 1, \ \text{and} \ F_S(-\infty) = 0 
\end{equation}
Then, \eq{\ref{eq:quantile_regression-solution}} can be derived as:
\begin{equation}
\begin{aligned}
    \frac{K}{\vert\itemSet\vert} F_S(\beta_u) =& \Big(1 - \frac{K}{\vert\itemSet\vert}\Big) \Big(1 - F_S(\beta_u)\Big) \\
    F_{S}(\beta_u) =& \Big(1 - \frac{K}{\vert\itemSet\vert}\Big)
\end{aligned}
\end{equation}

resulting $\int_{\beta_u}^{\infty} \diff F_S(S) = K/\vert\itemSet\vert$, \ie the optimal $\hat \beta_u$ is precisely the \topk quantile of scores $S$.
\end{proof}

However, directly utilize \eq{\ref{eqs:valina-quantile-regression}} requires to traverse over the complete item space. In recommendation scenarios, the complete item space is huge, and computational intensive. To address this issue, we propose the efficient quantile estimation in \eq{\ref{eqs:debias-quantile-loss}}, which supports sampling negative items to conduct unbiased \topk quantile estimation. \Cref{table:computational-efficiency} shows the training time cost in one epoch, indicating our proposed efficient quantile estimation is computational efficient.

\subsection{Proof of Unbiased \eq{\ref{eqs:debias-quantile-loss}}}\label[appendix]{appendix:quantile-regression}

Quantile regression method \citep{koenker2005quantile,hao2007quantile} is uniformly used for sample quantile estimation. Typically, the \topk quantile is estimated by the following loss:
\begin{equation}\label{eqs:valina-quantile-regression}
    \loss_{\text{QR}}(u) = \frac{1}{\vert \itemSet \vert} \sum\limits_{i\in\itemSet} \Big((1-\frac{K}{\vert\itemSet\vert})(\sui-\hat\beta_u)_+ + \frac{K}{\vert\itemSet\vert}(\hat\beta_u-\sui)_+\Big)
\end{equation}
With expectation $\E_{\sampledNegative}\left[ \sum_{j\in\sampledNegative}\rho_K(\suj-\hat{\beta}_{u})\right] = (\vert\sampledNegative\vert)/(\vert\itemSet\vert - \vert\posU\vert)$
$\sum_{j\in\negU}\rho_K(\suj-\hat{\beta}_{u})$, \eq{\ref{eqs:debias-quantile-loss}} can be derived as:
\[
\E_{\sampledNegative}[\mathcal{Q}_K(u)] \!\!=\!\! \frac{1}{\vert\itemSet\vert}\left(\sumPosU\!\!\rho_K(\sui-\hat{\beta}_{u}) + \!\!\sumNegU\!\!\rho_K(\suj-\hat{\beta}_{u})\right) = \loss_{\text{QR}}(u)  
\]

% table: optimal hyperparameters of IID setting
\section{Optimal Hyperparameters}

We report the optimal hyperparameters of each method on each dataset and backbone from \Cref{tables:OOD-hyperparameter,table:IID-Hyperparameter-official}, in the order of the hyperparameters listed in \Cref{tab:hyperparameters-searching}.

\begin{table}[h]
    \centering
    \caption{Hyperparameters to be searched for each method.}
    \begin{tabular}{l|l}
        \Xhline{1.3pt}
        \multicolumn{1}{c|}{\textbf{Method}} & \multicolumn{1}{c}{\textbf{Other Hyperparameters}} \bigstrut\\
        \Xhline{1pt}
        BPR         & no other hyperparameters                                                  \bigstrut[t]\\
        AATP        &  \{$\text{lr}_{\text{quantile}}$\}                                        \\
        RS@$K$      &  \{$\tau_{\text{inner}}$, $\tau_{\text{outside}}$\}                       \\
        SmoothI@$K$     &  \{$\tau$, $\delta$\}                                                 \\
        LLPAUC      & \{$\alpha$, $\beta$\}                                                     \\
            SL          &  \{$\tau$\}                                                           \\
        AdvInfoNCE  & \{$T_{\text{adv}}$, $E_{\text{adv}}$, $\text{lr}_{\text{adv}}$, $\tau$\}  \\
        BSL         &  \{$\tau_1$, $\tau_2$\}                                                   \\
        PSL          &  \{$\tau$\}                                                              \\
        \lossName   &  \{$\tau$\}                                                               \bigstrut[b]\\
        \Xhline{1.3pt}
    \end{tabular}
        \label{tab:hyperparameters-searching}
\end{table}

\begin{table}[htbp] 
    \scriptsize
    \centering
    \caption{Optimal hyperparameters of OOD setting.} 
    \begin{tabular}{c|l|cccc} 
    \Xhline{1.2pt} 
    \multicolumn{6}{c}{\textbf{Gowalla}} \bigstrut\\ 
    \Xhline{1.0pt}
    \textbf{Model} & \textbf{Loss} & \textbf{lr} & \textbf{wd} & \textbf{wl} & \textbf{others} \bigstrut\\ 
    \Xhline{1.0pt} 
    \multirow{10}[2]{*}{MF} 
    & BPR           & $10^{-3}$ & $10^{-8}$ & N/A & \bigstrut[t]\\ 
    & AATP          & $10^{-3}$ & $10^{-6}$ & N/A & \{$10^{-4}$\} \\ 
    & RS@$K$        & $10^{-2}$ & $10^{-8}$ & N/A & \{0.05, 5\} \\ 
    & SmoothI@$K$   & $10^{-3}$ & 0         & N/A & \{0.1, 0.1\} \\ 
    & SL            & $10^{-1}$ & 0         & N/A & \{0.1\} \\ 
    & BSL           & $10^{-1}$ & 0         & N/A & \{0.1, 0.1\} \\ 
    & PSL           & $10^{-1}$ & 0         & N/A & \{0.05\} \\ 
    & AdvInfoNCE    & $10^{-1}$ & 0         & N/A & \{5, 30, $10^{-5}$, 0.1\} \\ 
    & LLPAUC        & $10^{-3}$ & $10^{-6}$ & N/A & \{0.7, 0.01\} \\ 
    & TL@$K$        & $10^{-1}$ & 0         & N/A & \{0.05\} \\ 
    \Xhline{1.2pt} 
    \multicolumn{6}{c}{\textbf{Games}} \bigstrut\\ 
    \Xhline{1.0pt}
    \textbf{Model} & \textbf{Loss} & \textbf{lr} & \textbf{wd} & \textbf{wl} & \textbf{others} \bigstrut\\ 
    \Xhline{1.0pt} 
    \multirow{10}[2]{*}{MF} 
    & BPR           & $10^{-3}$ & $10^{-6}$ & N/A & \bigstrut[t]\\ 
    & AATP          & $10^{-3}$ & $10^{-6}$ & N/A & \{$10^{-2}$\} \\ 
    & RS@$K$        & $10^{-1}$ & $10^{-8}$ & N/A & \{0.05, 5\} \\ 
    & SmoothI@$K$   & $10^{-1}$ & 0         & N/A & \{0.1, 0.05\} \\ 
    & SL            & $10^{-2}$ & 0         & N/A & \{0.2\} \\ 
    & BSL           & $10^{-2}$ & 0         & N/A & \{0.1, 0.2\} \\ 
    & PSL           & $10^{-2}$ & $10^{-8}$ & N/A & \{0.1\} \\ 
    & AdvInfoNCE    & $10^{-2}$ & 0         & N/A & \{5, 5, $10^{-5}$, 0.2\} \\ 
    & LLPAUC        & $10^{-3}$ & $10^{-6}$ & N/A & \{0.6, 0.7\} \\ 
    & TL@$K$        & $10^{-2}$ & $10^{-6}$ & N/A & \{0.1\} \\ 
    \Xhline{1.2pt} 
    \end{tabular} 
    \label{tables:OOD-hyperparameter} 
\end{table}

% table: optimal hyperparameters of IID setting
\begin{table*}[t]
    \scriptsize
    \centering
    \caption{Optimal hyperparameters of IID setting.}
    \begin{tabular}{c|l|cccc|cccc}
    \Xhline{1.2pt}
    \multirow{2}[4]{*}{\textbf{Model}} & \multicolumn{1}{c|}{\multirow{2}[4]{*}{\textbf{Loss}}} & \multicolumn{4}{c|}{\textbf{Gowalla}} & \multicolumn{4}{c}{\textbf{Beauty}} \bigstrut\\
    \cline{3-10}          &       & \textbf{lr} & \textbf{wd} & \textbf{wl} & \textbf{others} & \textbf{lr} & \textbf{wd} & \textbf{wl}  & \textbf{others} \bigstrut\\
    \Xhline{1.0pt}
    \multirow{9}[2]{*}{MF} 
    & BPR           & $10^{-3}$ & $10^{-8}$ &   N/A     &               & $10^{-3}$ & $10^{-6}$     &   N/A     &   \bigstrut[t]\\
    & AATP          & $10^{-3}$ &  $10^{-6}$        &   N/A     & \{$10^{-4}$\}       & $10^{-2}$ & 0             &   N/A     &   \{$10^{-3}$\}      \\
    & RS@$K$        & $10^{-1}$ & 0         &   N/A     & \{0.01, 2\}       & $10^{-1}$ & $10^{-8}$             &   N/A     &   \{0.01, 1\}      \\
    & SmoothI@$K$        & $10^{-3}$ & 0         &   N/A     & \{0.1, 0.05\}       & $10^{-1}$ & $10^{-8}$             &   N/A     &   \{0.2, 0.05\}      \\
    & SL            & $10^{-1}$ & 0         &   N/A     & \{0.1\}       & $10^{-3}$ & $10^{-8}$             &   N/A     &   \{0.15\}      \\
    & BSL           & $10^{-1}$ & 0         &   N/A     & \{0.1, 0.1\}       & $10^{-1}$ & 0             &   N/A     &   \{0.05, 0.15\}      \\
    & PSL           & $10^{-1}$ & 0         &   N/A     & \{0.05\}       & $10^{-1}$ & 0             &   N/A     &   \{0.1\}      \\
    & AdvInfoNCE    & $10^{-1}$ & 0         &   N/A     & \{15, 25, $10^{-5}$, 0.1\}       & $10^{-1}$ & 0             &   N/A     &  \{15, 30, $10^{-5}$, 0.15\}     \\
    & LLPAUC        & $10^{-3}$ & $10^{-6}$         &   N/A     & \{0.8, 0.01\}       & $10^{-2}$ & $10^{-6}$             &   N/A     &   \{0.6, 0.9\}      \\
    & \lossName        & $10^{-2}$ & 0         &   N/A     & \{0.05\}       & $10^{-1}$ & $10^{-8}$             &   N/A     &   \{0.1\}      \\
    \Xhline{1.0pt}
    \multirow{9}[2]{*}{LGCN} 
    & BPR           & $10^{-3}$ & $10^{-8}$ &   N/A     &               & $10^{-3}$ & $10^{-6}$     &   N/A     &   \bigstrut[t]\\
    & AATP          & $10^{-3}$ & 0         &   N/A     & \{$10^{-2}$\}       & $10^{-3}$ & $10^{-8}$             &   N/A     &   \{$10^{-4}$\}      \\
    & RS@$K$        & $10^{-2}$ & 0         &   N/A     & \{0.1, 5\}       & $10^{-3}$ & 0             &   N/A     &   \{0.1, 5\}      \\
    & SmoothI@$K$        & $10^{-3}$ & $10^{-8}$         &   N/A     & \{0.1, 0.05\}       & $10^{-3}$ & $10^{-8}$             &   N/A     &   \{0.2, 0.05\}      \\
    & SL            & $10^{-1}$ & 0         &   N/A     & \{0.1\}       & $10^{-1}$ & 0             &   N/A     &   \{0.2\}      \\
    & BSL           & $10^{-1}$ & 0         &   N/A     & \{0.05, 0.1\}       & $10^{-1}$ & 0             &   N/A     &   \{0.15, 0.2\}      \\
    & PSL           & $10^{-2}$ & 0         &   N/A     & \{0.05\}       & $10^{-1}$ & 0             &   N/A     &   \{0.1\}      \\
    & AdvInfoNCE    & $10^{-2}$ & 0         &   N/A     & \{5, 30, $10^{-5}$, 0.1\}       & $10^{-1}$ & 0             &   N/A     &  \{5, 25, $10^{-5}$, 0.2\}     \\
    & LLPAUC        & $10^{-3}$ & $10^{-8}$         &   N/A     & \{0.1, 0.05\}       & $10^{-2}$ & $10^{-6}$             &   N/A     &   \{0.1, 0.1\}      \\
    & \lossName         & $10^{-2}$ & 0         &   N/A     & \{0.05\}       & $10^{-1}$ & 0             &   N/A     &   \{0.1\}      \\
    \Xhline{1.0pt}
    \multirow{9}[2]{*}{XSimGCL} 
    & BPR           & $10^{-3}$ & 0 &  0.05     &               & $10^{-2}$ & $10^{-4}$     &   0.2     &   \bigstrut[t]\\
    & AATP          & $10^{-4}$ & $10^{-6}$         &   0.05     & \{$10^{-4}$\}       & $10^{-2}$ & $10^{-8}$             &   0.05     &   \{$10^{-3}$\}      \\
    & RS@$K$        & $10^{-2}$ & 0         &   0.01     & \{0.1, 5\}       & $10^{-3}$ & 0             &   0.01     &   \{0.15, 5\}      \\
    & SmoothI@$K$        & $10^{-2}$ & 0         &   0.01     & \{0.1, 0.1\}       & $10^{-2}$ & 0             &   0.01     &   \{0.1, 0.05\}      \\
    & SL            & $10^{-3}$ & 0         &   0.01     & \{0.1\}       & $10^{-3}$ & 0             &   0.01     &   \{0.2\}      \\
    & BSL           & $10^{-2}$ & 0         &   0.01     & \{0.05, 0.1\}       & $10^{-3}$ & $10^{-8}$             &   0.01     &   \{0.15, 0.02\}      \\
    & PSL           & $10^{-1}$ & 0         &   0.01     & \{0.05\}       & $10^{-2}$ & 0             &   0.01     &   \{0.1\}      \\
    & AdvInfoNCE    & $10^{-3}$ & 0         &   0.01     & \{15, 30, $10^{-5}$, 0.1\}       & $10^{-3}$ & 0             &   0.01     &  \{10, 20, $10^{-5}$, 0.2\}     \\
    & LLPAUC        & $10^{-3}$ & $10^{-8}$         &   0.01     & \{0.1, 0.01\}       & $10^{-2}$ & $10^{-6}$             &   0.01     &   \{0.1, 0.01\}      \\
    & \lossName         & $10^{-1}$ & 0         &   0.01     & \{0.05\}       & $10^{-1}$ & 0             &   0.01     &   \{0.1\}      \\
    \Xhline{1.2pt}
    \multirow{2}[4]{*}{\textbf{Model}} & \multicolumn{1}{c|}{\multirow{2}[4]{*}{\textbf{Loss}}} & \multicolumn{4}{c|}{\textbf{Games}} & \multicolumn{4}{c}{\textbf{Electronics}} \bigstrut\\
    \cline{3-10}          &       & \textbf{lr} & \textbf{wd} & \textbf{wl} & \textbf{others} & \textbf{lr} & \textbf{wd} & \textbf{wl}  & \textbf{others} \bigstrut\\
    \Xhline{1.0pt}
    \multirow{9}[2]{*}{MF} 
    & BPR           & $10^{-2}$ & $10^{-6}$ &   N/A     &               & $10^{-3}$ & $10^{-6}$     &   N/A     &   \bigstrut[t]\\
    & AATP          & $10^{-3}$ & $10^{-6}$         &   N/A     & \{$10^{-2}$\}       & $10^{-2}$ & 0             &   N/A     &   \{$10^{-4}$\}      \\
    & RS@$K$        & $10^{-1}$ & $10^{-8}$         &   N/A     & \{0.05, 5\}       & $10^{-3}$ & $10^{-6}$             &   N/A     &   \{0.05, 5\}      \\
    & SmoothI@$K$        & $10^{-2}$ & 0         &   N/A     & \{0.15, 0.05\}       & $10^{-3}$ & $10^{-8}$             &   N/A     &   \{0.15, 0.2\}      \\
    & SL            & $10^{-2}$ & $10^{-6}$         &   N/A     & \{0.2\}       & $10^{-3}$ & $10^{-6}$             &   N/A     &   \{0.2\}      \\
    & BSL           & $10^{-1}$ & $0$         &   N/A     & \{0.05, 0.2\}       & $10^{-2}$ & $10^{-8}$             &   N/A     &   \{0.1, 0.2\}      \\
    & PSL           & $10^{-1}$ & 0         &   N/A     & \{0.1\}       & $10^{-3}$ & $10^{-8}$             &   N/A     &   \{0.1\}      \\
    & AdvInfoNCE    & $10^{-1}$ & $10^{-8}$         &   N/A     & \{15, 5, $10^{-5}$, 0.2\}       & $10^{-2}$ & $10^{-8}$             &   N/A     &  \{10, 30, $10^{-5}$, 0.15\}     \\
    & LLPAUC        & $10^{-2}$ & $10^{-6}$         &   N/A     & \{0.5, 0.9\}       & $10^{-3}$ & $10^{-6}$             &   N/A     &   \{0.6, 0.05\}      \\
    & \lossName         & $10^{-1}$ & 0         &   N/A     & \{0.1\}       & $10^{-3}$ & $10^{-6}$             &   N/A     &   \{0.1\}      \\
    \Xhline{1.0pt}
    \multirow{9}[2]{*}{LGCN} 
    & BPR           & $10^{-3}$ & $10^{-6}$ &   N/A     &               & $10^{-3}$ & $10^{-6}$     &   N/A     &   \bigstrut[t]\\
    & AATP          & $10^{-1}$ & 0         &   N/A     & \{$10^{-2}$\}       & $10^{-1}$ & 0             &   N/A     &   \{$10^{-2}$\}      \\
    & RS@$K$        & $10^{-1}$ & 0         &   N/A     & \{0.1, 5\}       & $10^{-3}$ & $10^{-6}$             &   N/A     &   \{0.15, 4\}      \\
    & SmoothI@$K$        & $10^{-3}$ & $10^{-8}$         &   N/A     & \{0.15, 0.05\}       & $10^{-3}$ & $10^{-8}$             &   N/A     &   \{0.15, 0.1\}      \\
    & SL            & $10^{-1}$ & 0         &   N/A     & \{0.2\}       & $10^{-2}$ & 0             &   N/A     &   \{0.2\}      \\
    & BSL           & $10^{-1}$ & $10^{-8}$         &   N/A     & \{0.05, 0.2\}       & $10^{-2}$ & 0             &   N/A     &   \{0.1, 0.2\}      \\
    & PSL           & $10^{-1}$ & 0         &   N/A     & \{0.1\}       & $10^{-3}$ & $10^{-8}$             &   N/A     &   \{0.1\}      \\
    & AdvInfoNCE    & $10^{-1}$ & 0         &   N/A     & \{20, 15, $10^{-5}$, 0.2\}       & $10^{-2}$ & 0             &   N/A     &  \{15, 30, $10^{-5}$, 0.2\}     \\
    & LLPAUC        & $10^{-2}$ & $10^{-6}$         &   N/A     & \{0.1, 0.1\}       & $10^{-3}$ & $10^{-6}$             &   N/A     &   \{0.5, 0.01\}      \\
    & \lossName         & $10^{-1}$ & $0$         &   N/A     & \{0.1\}       & $10^{-1}$ & 0             &   N/A     &   \{0.1\}      \\
    \Xhline{1.0pt}
    \multirow{9}[2]{*}{XSimGCL} 
    & BPR           & $10^{-3}$ & $10^{-6}$ &   0.1     &               & $10^{-3}$ & $10^{-5}$     &   0.2     &   \bigstrut[t]\\
    & AATP          & $10^{-2}$ & $10^{-8}$         &   0.01     & \{$10^{-2}$\}       & $10^{-3}$ & $10^{-8}$             &   0.01     &   \{$10^{-2}$\}      \\
    & RS@$K$        & $10^{-3}$ & $10^{-8}$         &   0.01     & \{0.15, 5\}       & $10^{-2}$ & $10^{-6}$             &   0.01     &   \{0.1, 4\}      \\
    & SmoothI@$K$        & $10^{-1}$ & 0         &   0.01     & \{0.15, 0.05\}       & $10^{-3}$ & $10^{-8}$             &  0.01     &   \{0.1, 0.1\}      \\
    & SL            & $10^{-2}$ & 0         &   0.01     & \{0.2\}       & $10^{-1}$ & 0             &   0.01    &   \{0.2\}      \\
    & BSL           & $10^{-2}$ & 0         &  0.01    & \{0.05, 0.2\}       & $10^{-2}$ & 0             &   0.01   &   \{0.25, 0.25\}      \\
    & PSL           & $10^{-2}$ & 0         &   0.01   & \{0.1\}       & $10^{-2}$ & 0             &   0.01  &   \{0.1\}      \\
    & AdvInfoNCE    & $10^{-2}$ & 0         &  0.01   & \{15, 20, $10^{-4}$, 0.2\}       & $10^{-3}$ & 0             & 0.01    &  \{15, 30, $10^{-5}$, 0.2\}     \\
    & LLPAUC        & $10^{-3}$ & $10^{-6}$         &  0.01  & \{0.6, 0.05\}       & $10^{-2}$ & $10^{-6}$             &  0.05   &   \{0.7, 0.01\}      \\
    & \lossName        & $10^{-2}$ & $10^{-8}$         &  0.01   & \{0.1\}       & $10^{-1}$ & 0             &  0.01&   \{0.1\}      \\
    \Xhline{1.2pt}
    \end{tabular}
                \label{table:IID-Hyperparameter-official}
\end{table*}

\end{document}